\documentclass[]{elsart}
\usepackage{natbib}
\usepackage{graphicx}
\usepackage{amsmath,amssymb}
\usepackage{color}		
\renewcommand{\cite}{\citep}

\newcommand*{\ie}[1][]{{\it i.e.}\ }
\newcommand*{\eg}[1][]{{\it e.g.}\,}
\newcommand*{\etc}[1][]{{\it etc.}\,}
\newcommand*{\const}[1][]{{\it const}\,}
\newcommand*{\Jc}[1][]{\mbox{$\mathcal{J}$}}
\newcommand*{\Ld}[1][]{\mbox{$\mathcal{L}$}}

\newcommand{\V}{\mbox{$\mathcal{V}$}}
\newcommand*{\R}[1][]{\mbox{$\mathcal{R}$}}
\newcommand*{\dd}[1][]{\mbox{\textrm{d}}}
\newcommand*{\DD}[1][]{\mbox{\textrm{D}}}
\newcommand*{\pl}[1][]{\partial}
\newcommand{\bnabla}{\mbox{\boldmath $\nabla$}}

\newcommand{\grad}{\mbox{grad}}

\newcommand{\surf}{\mbox{\boldmath $s$}^2 }
\newcommand{\Surf}{\mbox{\boldmath $S$}^2 }

\newcommand*{\al}[1][]{\alpha}
\newcommand*{\ga}[1][]{\gamma}
\newcommand*{\Del}[1][]{\Delta}
\newcommand*{\del}[1][]{\delta}
\newcommand{\eps}{\epsilon}
\newcommand*{\vep}[1][]{\varepsilon}
\newcommand*{\om}[1][]{\omega}
\newcommand*{\Om}[1][]{\Omega}
\newcommand*{\Th}[1][]{\Theta}

\newcommand*{\bom}[1][]{\mbox{\boldmath $\omega$}}
\newcommand*{\bOm}[1][]{\mbox{\boldmath $\Omega$}}
\newcommand*{\bxi}[1][]{\mbox{\boldmath $\xi$}}
\newcommand{\bta}{\mbox{\boldmath $\eta$}}
\newcommand{\bPsi}{\mbox{\boldmath $\Psi$}}

\newcommand{\ba}{\mbox{\boldmath $a$}}
\newcommand*{\A}[1][]{\mbox{$\mathcal{A}$}}
\newcommand*{\bA}[1][]{\mbox{\boldmath $A$}}
\newcommand*{\bb}[1][]{\mbox{\boldmath $b$}}
\newcommand*{\bB}[1][]{\mbox{\boldmath $B$}}
\newcommand*{\bE}[1][]{\mbox{\boldmath $E$}}
\newcommand*{\bF}[1][]{\mbox{\boldmath $F$}}
\newcommand*{\G}[1][]{\mbox{$\mathcal{G}$}}
\newcommand*{\bU}[1][]{\mbox{\boldmath $U$}}
\newcommand*{\bn}[1][]{\mbox{\boldmath $n$}}
\newcommand*{\bp}[1][]{\mbox{\boldmath $p$}}
\newcommand*{\bQ}[1][]{\mbox{\boldmath $Q$}}
\newcommand*{\bs}[1][]{\mbox{\boldmath $s$}}

\newcommand*{\bv}[1][]{\mbox{\boldmath $v$}}
\newcommand*{\bV}[1][]{\mbox{\boldmath $V$}}
\newcommand*{\bw}[1][]{\mbox{\boldmath $w$}}
\newcommand*{\bx}[1][]{\mbox{\boldmath $x$}}
\newcommand{\bX}{\mbox{\boldmath $X$}}

\begin{document}
\vspace*{-10mm}  
{\small {\it Fluid Dynamics Research} (2008), accepted on 1 December 2007. }

\vspace*{10mm}

\begin{frontmatter}
\title{Variational formulation of ideal fluid flows according to gauge principle}
\author{Tsutomu  Kambe\corauthref{cor1}}
\corauth[cor1]{kambe@ruby.dti.ne.jp, \ http://www.purple.dti.ne.jp/kambe/}
\address{IDS, Higashi-yama 2-11-3, Meguro-ku, Tokyo 153-0043, Japan }
{\footnotesize 
\begin{center}
Received 6 February 2007; revised 23 November 2007;  accepted 1 December 2007 \\
                        Communicated by Y. Fukumoto.  
\end{center} }

\begin{abstract}
On the basis of the gauge principle of field theory, a new variational formulation 
is presented for flows of an ideal fluid.  The fluid is defined thermodynamically 
by mass density and entropy density, and its flow fields  are characterized by 
symmetries of translation and rotation.  The rotational transformations are regarded as 
gauge transformations as well as the translational ones.   In addition to the Lagrangians 
representing the translation symmetry, a structure of  rotation symmetry is 
equipped with a Lagrangian $\Lambda_A$ including the vorticity and a vector potential
bilinearly. Euler's equation of motion is derived from variations according to the 
action principle. In addition, the equations of continuity and  entropy are derived  
from the variations.  Equations of conserved currents are deduced as the Noether 
theorem in the space of Lagrangian  coordinate $\ba$.  Without $\Lambda_A$, the action 
principle results in the Clebsch solution with vanishing helicity.  The Lagrangian 
$\Lambda_A$ yields non-vanishing vorticity and provides a source term of non-vanishing 
helicity. The vorticity equation is derived as an equation of the gauge field, and 
the $\Lambda_A$ characterizes  topology of the field.  The present formulation 
is comprehensive and provides a consistent basis for a unique transformation between 
the Lagrangian $\ba$ space and the Eulerian $\bx$ space.  In contrast, with translation 
symmetry alone, there is an arbitrariness in the transformation between these spaces.
\end{abstract}

\begin{keyword}
Gauge priciple, Variational formulation, Ideal fluid, Rotation symmetry,
Chern-Simons term
\end{keyword}

\end{frontmatter}

\newpage
\section{Introduction \label{S1-Int}}

A guiding principle in physics is that physical laws should be expressed in  a form 
that is independent of any particular coordinate system.  {\it Fluid mechanics} is 
a field theory of Newtonian mechanics that has Galilean symmetry: that is to say, it 
is covariant under transformations of the Galilei group.  Two symmetries (\ie 
transformation invariances) are known as subgroups of the Galilei group: 
translation (space and time) and space-rotation. In the present formulation of 
flows of an ideal fluid, we seek a scenario which has a formal equivalence with the 
gauge theory in physics.  The gauge theory provides a basis for reflection on the 
similarity between fluid mechanics and other physical fields. 

The gauge principle\footnote{See Appendix, or standard texts such as Weinberg (1995), 
Frankel (1997) or Aitchison \& Hey (1982).} requires a physical system under 
investigation to have a {\it symmetry}, \ie a gauge invariance (an invariance with 
respect to a certain group of transformations).  Following this principle, an attempt 
has been made in Kambe (2007) to study a gauge symmetry of flow fields with respect 
to translational transformations.  The formulation started from a Galilei-invariant 
Lagrangian of a system of {\it point masses} which is known to have {\it global} gauge 
symmetries with respect to both translation and rotation.\footnote{These symmetries 
are interpreted as the {\it homegeneity} and {\it isotropy} of space in \S7 and 9 of 
Landau \& Lifshitz (1976).  See \S\ref{S2-1} and Appendix for the definition of global  
gauge symmetry (\ie global gauge invariance) and local gauge symmetry.}
It was then extended to flows of a fluid, a {\it continuous} material characterized with 
mass density and entropy density. In addition to the global symmetry, {\it local} 
gauge invariance of a Lagrangian is required for such a continuous field. Thus, the 
convective derivative of fluid mechanics (\ie the Lagrange derivative) is identified 
as a {\it covariant derivative}, which is a building block in the framework of gauge 
theory.  Based on this, appropriate Lagrangian functionals are defined  for motion of 
an ideal fluid.  Euler's equation of motion is derived from the action principle. 
It is found that a general solution of this formulation is equivalent to the classical 
Clebsch solution (Clebsch 1859; Lamb 1932), in which the vorticity has a particular form
and the helicity vanishes.

In most traditional formulations, the continuity equation and  isentropic condition 
are taken into account as constraints for the variations of the action principle 
by using Lagrange multipliers, where the {\it isentropy} means that each fluid particle 
of an ideal fluid keeps its entropy value along its trajectory (but the fluid is 
not necessarily {\it homentropic}\footnote{A fluid is {\it homentropic} if its 
entropy per unit mass is uniform throughout space.}.)  In the present formulation 
however, those equations are derived from the action principle. The total Lagrangian 
density consists of kinetic energy and internal energy (with negative sign), 
supplemented with additional two terms associated with translation invariances of 
mass and entropy (see (27) of \S3.1). As is interpreted in detail in \S4.1, it is 
remarkable that  the Lagrangians associated with the latter two terms are given 
such forms as can be eliminated from the action represented with the  particle 
coordinate $\ba$ (\ie Lagrangian coordinate),  so that the Euler-Lagrange equations 
represented by the independent varibales $\ba=(a^1,a^2,a^3)$ are uninfluenced by the 
additional terms.  Such a form of Lagrangian has close analogy to the Chern-Simons 
term in the gauge theory.

Earlier papers (Kambe, 2003a, b) investigated the rotation symmetry of the velocity 
field $\bv(\bx)$ in a preliminary way, and found  that the vorticity 
$\bom=\nabla\times \bv$ is in fact a gauge field associated with the rotation symmetry.  
However in the previous studies including Kambe (2007), the vorticity and helicity 
were not taken into account in satisfactory manner. Present formulation tries to 
account for the vorticity and helicity more properly and  comprehensively.  Thus, 
the main theme of the present paper is the rotation symmetry. 

A new structure of the rotation symmetry is given in \S6 (after a review of the 
rotational transformations in \S5) by the following Lagrangian $\Lambda_A$:
\[	\Lambda_A =  \int_{M} \langle \bA,\, E_W[\bom] \,\rangle \,\dd^3\bx,  \]  
where $\bA$ is a vector potential and $E_W[\bom] \equiv \pl_t \bom 
+(\bv\cdot\nabla)\bom -(\bom\cdot\nabla)\bv + (\nabla\cdot \bv) \bom$.  This is derived 
from a Lagrangian form characteristic of a {\it topological} term known in the gauge 
theory.  This new term yields non-vanishing rotational component of the velocity field, 
and provides a source term of non-vanishing helicity. This is also the 
Chern-Simons term, describing non-trivial topology of vorticity field, \ie mutual 
linking of vorticity lines.  The vorticity equation is derived as an equation for 
the gauge field. These are described in \S\ref{S6} (Lagrangian $L_A$ associated with 
rotation symmetry),  \S\ref{S7} (Corrections to the Variation), and \S\ref{S8}
(Significance  of  $\Lambda_A$).

With regard to the variational formulation of fluid flows, the works of Eckart 
(1938, 1960) and Herivel (1955) are among the earliest to have influenced current 
formulations.  Their variations are carried out in two ways: \ie a Lagrangian 
approach and an Eulerian approach.  In both approaches, the equation of continuity 
and the condition of is entropy are added as constraint conditions on the variations 
by means of Lagrange multipliers.  The Lagrangian approach is also taken by Soper 
(1976). In this relativistic formulation those equation  are derived from the 
equations of current conservation.  Several action principles to describe relativistic 
fluid dynamics have appeared in the past (see Soper (1976, \S4.2) for some list of them), 
not only that of Taub (1954).

Firstly, in the  Lagrangian approach,  the Euler-Lagrange equation results in an equation 
equivalent to Euler's equation of motion in which the acceleration term is represented 
as the second time derivative of position coordinates of the Lagrangian representation.  
In this formulation, however, there is a certain degree of freedom in the relation 
between the Lagrangian particle coordinates $(a^1,a^2,a^3)$ and Eulerian space 
coordinates $(x^1,x^2,x^3)$. Namely, the relation between them is determined only up 
to an unknown rotation. This point has not been considered seriously in the past, 
although it is fundamental for the formualtion of flows.  This will be discussed more 
precisely in the sections 4.2 and 8.2. A common feature of Herivel (1955) and Eckart 
(1960) is that both arrive at the Clebsch solution in the Eulerian representaion (see 
Sec.\ref{A1-GI} and \ref{3B}).  However, it was remarked by Bretherton (1970) that 
the Clebsch representation has only local validity in the neighbourhood of a chosen 
point  if vortex lines are knotted or linked.

In the second approach, referred to as the {\it Eulerian} description, the action 
principle of an ideal fluid results only in potential flows if the fluid is 
homentropic, as shown in Kambe (2003a, 2007).  However it is well-known that, even 
in such a homentropic fluid, it should be possible to have rotational 
flows.\footnote{In fact, Euler (1755) showed that his equation of motion can drescribe 
rotational flows, in the historical paper: L.~Euler, {\it Principes g\'{e}n\'{e}raux 
du mouvement des fluides}, MASB, v.\,{\bf 11} (1757), 274-315. (MASB: Memoires de 
l'Acad\'{e}mie Ryale des Sciences, ann\'{e}e 1755. Berlin 1757.)}  
This is a long-standing problem (Serrin 1959; Lin 1963; Seliger \& Whitham 1968; 
Bretherton 1970; Salmon 1988).  Lin (1963) tried to resolve this difficulty by 
introducing a {\it constraint} as a side condition, imposing invariance  of Lagrangian 
particle labels $(a^1,a^2,a^3)$ along particle trajectories.  The constraints for 
the variation are defined by using Lagrange multipliers (functions of positions), 
which are called 'potentials'.  In addition, the continuity equation and  isentropic 
condition are also taken into account by using Lagrange multipliers.  However, the 
physical significance of those potentials introduced as the Lagrange multipliers 
is not clear.  Mysteriously, the Lagrange multiplier for the continuity equation 
becomes  the velocity potential for the irrotational part of the velocity.
Gauge theory for fluid flows  provides  a crucial key to resolve these issues, which is
the main target of the present investigation.  

Symmetries imply conservation laws. According to the relativistic formulation 
(\eg \,Soper 1976; Landau \& Lifshitz 1987 (Ch.XV)), as long as the flow fields obey 
the Euler-Lagrange equation, an energy-momentum tensor $T^{\mu\nu}$ must obey the 
conservation equation of the form $\pl_\nu\, T^{\mu\nu}=0$ (considered in 
Sec.~\ref{4B} below).  The Energy-momentum tensor $T^{\mu\nu}$ in the 
present case is given in Sec.\ref{4-FLV}.  A similar field-theoretic approach is 
taken in Jackiw (2002) by applying the ideas of particle physics to fluid mechanics 
in terms of Hamiltonian and  Poisson brackets, both relativistically and 
nonrelativistically, and extension to supersymmetry is also considered.  
In Soper (1976) and Jackiw (2002), the  nonrelativistic part follows the traditional 
approach and no explicit gauge-theoretic consideration is given to fluid mechanics 
in the sense of the present formulation. 

The appendix describes the background of the present theory: (A) Known gauge invariances
in two systems of electromagnetism and Clebsch form of fluid flows, (B) Related aspects 
in quantum mechanics and relativistic theory, and (C) Brief scenario of gauge principle.

\section{Translation symmetry of flow fields  \  \label{S2-TS} }

The formulation of Kambe (2007) started from a Galilei-invariant Lagrangian of 
a system of {\it point masses}, which is known to have {\it global} gauge symmetries 
with respect to both translation and rotation.  Concept of {\it local} transformation
is an extention of the {\it global} transformation to describe  a continuous field. 
In order to represent a continuous distribution of mass, the discrete positions of point 
masses are replaced by continuous parameters $\ba =(a^1, a^2, a^3)$  of particles in 
a bounded sub-space $M$ (of consideration) in the three-dimensional Euclidean space $E^3$.  
The spatial position of each mass particle labelled by $\ba$ (Lagrange parameter) is 
denoted by $\bx=X_a(t) \equiv X(\ba, t)$, a function of $\ba$ as well as the time $t$.  
An Eulerian space point is denoted by $\bx =(x^1, x^2, x^3)$.  The fluid  particle 
occupying the point $\bx$ at time $t$ is denoted by $\ba(\bx, t)$.\footnote{Here 
the Lagrangian coordinates $\ba=(a^1, a^2, a^3)$ are defined so as to represent 
the {\it mass coordinate} such that $\dd^3 \ba = \rho\,\dd^3\bx$, where 
$\dd^3 \ba= \dd a^1 \dd a^2 \dd a^3$ and $\dd^3 \bx= \dd x^1 \dd x^2 \dd x^3$. 
A variable such as $X_a$ denotes the one with respect to the particle $\ba$.} 

\subsection{Global invariance of a fluid \label{S2-1}}

For  a continuous distribution of mass ({\it i.e.} for a {\it fluid}),  Lagrangian  
functional is defined as 
\begin{equation} 
\Lambda = \int L(x, v) \ \dd^3 x  \,, \label{L-CF} 
\end{equation}
where $x$ and $v$ denote $X_a^k(t)$ and $v^k = \pl_t X_a^k(t)$ ($k=1,2,3$) respectively,
and $L(x,v)$ is a Lagrangian density.  Suppose that  an infinitesimal transformation is 
expressed by $x \to x'= x + \delta x$ and $v \to v'= v+ \delta v$, where
$\delta x = \xi(x,t)$ and $\delta v = \pl_t(\delta x)$, a component of $\xi$ is denoted 
by $\xi^k(x,t)$, an arbitrary differentiable variation field.  The 
resulting variation of the Lagrangian density $L(x, \,v)$ is 
\[  \delta L = \Big(\frac{\pl L}{\pl x^k} - \pl_t \big( \frac{\pl L}{\pl v^k} 
	\big) \Big)  \, \delta x^k + \pl_t \big(\frac{\pl L}{\pl v^k}\, 
	\delta x^k \big). 	\]
This  does not vanish in general. In fact, assuming the Euler-Lagrange equation,
$\pl L/\pl x^k - \pl_t (\pl L/\pl v^k ) =0$,  we obtain 
\begin{equation}
\delta L = \pl_t \big(\frac{\pl L}{\pl v^k}\, \delta x^k \big)
  =  \pl_t \big(\frac{\pl L}{\pl v^k} \big) \, \xi^k 
     +  \frac{\pl L}{\pl v^k}\, \pl_t \xi^k .  \label{dL3}
\end{equation}
Global invariance, \ie $\del \Lambda= \int \del L(x, v) \ \dd^3 x =0$
for arbitrary constants $\xi^k$ ($k=1,2,3$), requires that the total momentum defined by 
$\int (\pl L/\pl v^k)\,\dd^3 x$ must be invariant.  In the local transformation, 
however,  the variation field $\xi^k$ depends on  the time $t$ and space point $x$, 
and the variation $\del \Lambda= \int \delta L \,\dd^3 x$ does not vanish in general.

\subsection{Covariant derivative \label{2B-CD}}

According to the gauge principle of Appendix \ref{A3-GP}, non-vanishing of 
$\delta \Lambda$ is understood as meaning that  a {\it new field} $G$  must be taken 
into account in order to achieve {\it local gauge invariance} of the Lagrangian 
$\Lambda$.  To that end, we try to replace the partial time derivative $\pl_t$ by 
a {\it covariant derivative} $\DD_t$, where the derivative $\DD_t$ is defined by 
$\DD_t := \pl_t + G$.		
In dynamical systems like the present one, the {\it time derivative} is the primary 
object to be considered in the analysis of local gauge transformation, as implied by 
the expression (\ref{dL3}). Thus,  the time derivatives $\pl_t \xi$ and $\pl_t q$ 
are replaced by $\DD_t \xi = \pl_t \xi + G\,\xi$ and $\DD_t q = \pl_t q + G\,q$, 
where $q$ is understood to denote $\bx$.   A free-field Lagrangian $\Lambda_{\rm f}$ 
is defined by
\begin{equation}
\Lambda_{\rm f} = \int L_{\rm f}(v) \ \dd^3 \bx  :=  \half \int \langle v, \, v 
	\rangle \ \rho\, \dd^3 x = \half \int \langle \DD_t q, \,\DD_t q \rangle 
	\ \dd^3\ba  \,, \label{Lfvv} 
\end{equation}
where $ \rho \dd^3 x \equiv \dd^3 \ba$ denotes the mass element in a volume element 
$\dd^3 \bx$ of the $x$-space with $\rho$ the mass-density, \ and 
$v=\DD_t q$.

\subsection{Galilean transformation \label{S2-3} }

\subsubsection{Global  transformation  \label{2B-gGt}}

Symmetries to be investigated are the {\it translation symmetry} and {\it rotation 
symmetry},  which exist in the discrete system.  The Lagrangian 
$\Lambda_{\rm f}$  defined by (\ref{Lfvv}) for a continuous  field has such global 
symmetries, inherited from the discrete system.    It is a primary concern here  
to investigate whether the system of fluid flows satisfies local invariance. 
We consider the translation symmetry first, and in the second half of the 
present analysis we investigate the rotation symmetry.

A translational transformation from one frame $F$ to another $F'$ moving with 
a relative velocity $\bU$ is a Galilean transformation in Newtonian mechanics.
The transformation  is defined by 
\begin{equation}  
x \equiv (t,\ \bx) \ \ \Rightarrow \ \ x'\equiv (t',\ \bx')=(t,\ \bx - \bU\,t). 
\label{GTx}
\end{equation}
This is a sequence of global translations with a parameter $t$.
Corresponding transformation of velocity is  as follows:
$v \equiv (1,\ \bv) \ \Rightarrow \ v' \equiv (1, \ \bv')=(1,\ \bv - \bU)$. From 
(\ref{GTx}), differential operators $\pl_t = \pl/ \pl t$ and $\pl_k = \pl/ \pl x_k$ 
($\pl'_k = \pl/ \pl x'_k$) are transformed according to $\pl_t = \pl_{t'} 
- \bU\cdot \bnabla'$ and  $\bnabla = \bnabla'$, where $\bnabla=(\pl_1,\pl_2,\pl_3)$
and $\bnabla'=(\pl'_1,\pl'_2,\pl'_3)$.  There is a transformation invariance of the 
convective derivative $\DD_c \equiv \pl_t  + (\bv\cdot\bnabla)$: namely 
$\pl_t + (\bv\cdot\bnabla)=\pl_{t'} + (\bv'\cdot\bnabla')$.

\subsubsection{Local Galilean transformation \label{2C-lGt}}

The velocity field $\bv(\bx,t)$ is {\it defined} by the particle velocity, \ie 
\begin{equation}
\bv(X_a,t) = \frac{\dd }{\dd t} X_a(t).		\label{def-v}	
\end{equation}
Consider the following infinitesimal transformation:
\begin{equation}
 \bx'(\bx,t) = \bx + \bxi(\bx, t), \hskip10mm t' = t.    \label{lGtx}	
\end{equation}
This may be regarded as a local gauge transformation between non-inertial frames. 
In fact, the transformations (\ref{lGtx}) is understood to mean that the coordinate 
$\bx$ of a fluid particle at $\bx=X_a(t)$ in the frame $F$ 
is transformed to the new coordinate $\bx'$ of $F'$, given by 
$\bx'= X'_a(X_a,t) = X_a(t) + \bxi(X_a, t)$.  Therefore, its velocity 
$\bv=(\dd/\dd t)X_a(t)$  is transformed by
\begin{eqnarray}
  \bv' (\bx')  & \equiv &  \frac{\dd }{\dd t}X'_a  = \frac{\dd }{\dd t}\big( 
    X_a(t) + \bxi(X_a, t) \big) = \bv(X_a) + (\dd/\dd t) \bxi_a, \label{lGtu} \\
(\dd/\dd t) \bxi_a  & = & \pl_t\bxi + (\bv\cdot \nabla)\bxi\,\Big|_{\bx=X_a}, 
	\hskip10mm \bxi_a \equiv \bxi(X_a, t). 	\label{lGxi}
\end{eqnarray}
This is interpreted as follows: the origin of a local frame is displaced by 
$- \bxi$ where its coordinate axes  are moving  (without rotation)  with the 
velocity $- (\dd/\dd t) \bxi$ in accelerating motion (non-inertial frame). This implies 
that the velocity $\bv(\bx)$  is transformed according to (\ref{lGtu}), 
since the local frame is moving with the velocity 
$- (\dd/\dd t) \bxi$. Note that, in this transformation, the points 
$\bx$ and $\bx'$ are the same  with respect to the space $F$. 
In view of the transformation (\ref{lGtx}), the time derivative and spatial 
derivatives are transformed as 
\begin{eqnarray} 
\pl_t  & = & \pl_{t'} + (\pl_t\bxi) \cdot \nabla',  \hskip10mm 
		\nabla' = (\pl'_{\,k}),			\label{lGdt}	 \\
\pl_k  & = & \pl'_{\,k} + \ \pl_k \xi_l \ \pl'_{\,l}, \hskip12mm 
		\pl'_{\,k}  = \pl/\pl x'_k \,.    	\label{lGdx}
\end{eqnarray} 

\subsection{Invariance of $\DD_t$ and definition of velocity  \label{2E}}

In this local transformation too, there is an invariance of the convective derivative 
$\DD_c  \equiv \pl_t  + (\bv\cdot\bnabla) = \pl_{t'} + (\bv'\cdot\bnabla')$. 
In fact, from (\ref{lGdx}) and (\ref{lGtu}), we obtain
\[ \bv\cdot\bnabla = \bv\cdot\bnabla' + (\bv\cdot \bnabla \xi)\cdot \bnabla'
     = \bv'(\bx') \cdot\bnabla' + \big( - (\dd \bxi/\dd t) + \bv\cdot \bnabla \xi \big)
 	\cdot \nabla',		\]
where $\bv=\bv'- \dd \bxi/\dd t$.  The last term is $-\pl_t\bxi \cdot \nabla'$
by (\ref{lGxi}).  Thus, we find the invariance of $\DD_c$  by (\ref{lGdt}).
This invariance (\ie  the transformation symmetry) implies that the covariant 
derivative $\DD_t$ is in fact given by the convective derivative $\DD_c$: 
\begin{equation} 
\DD_t \equiv \pl_t + (\bv\cdot\bnabla) =\pl_{t'} + (\bv'\cdot\bnabla')  \label{CV-Dt}
\end{equation}
(see Kambe (2007) for its direct derivation).

For a scalar function $f(\bx,t)$, we have $\DD_t f = \pl_t f + \bv\cdot \nabla f$ 
(Lagrange derivative of $f(\bx,t)$). It is important to recognize that this derivative 
$\DD_t f$ has an intrinsic meaning. That is, we have the invariance: 
$\DD_t f = \DD'_{t'} f'$, with respect to the local Galilean transformation 
(\ref{lGtx}), since a scalar means $f'\equiv f'(\bx',t')= f(\bx,t)$.  In general,  
vectors or tensors need not have this property.

By using the covariant derivative $\DD_t$, we can define the velocity $\bv$ by 
$\DD_t \bx$.  In fact, each component $a^i$ of the particle label $\ba$ 
is a scalar, and satisfies $\DD_t a^i  = \pl_t a^i + (\bv\cdot \nabla)a^i =0$, 
since the particle with the label $a^i$ moves with the velocity $\bv=\pl_t X(\ba, t)$ 
by the definition  (\ref{def-v}). Setting  $\bx= X(\ba, t)$ for the particle position, 
we have 
\begin{equation}
\DD_t X(\ba, t) = \pl_t X(\ba, t) + \DD_t \ba \cdot 
	\nabla_a X = \pl_t X(\ba, t) = \bv ,			\label{DtX}
\end{equation}
where $(\nabla_a \bX)= (\pl X^k/\pl a^l)$.  In addition, applying $\DD_t$ to 
(\ref{lGtx}) and using  $\DD'_{t'}= \DD_t$, we have $\bv'= \DD'_{t'}\bx' 
= \DD_t (\bx + \bxi) = \bv + \DD_t\bxi$ (consistent with (\ref{lGtu})).   
Thus, the particle velocity $\bv$ can be defined by $\bv(\bx,t) = \DD_t \bx$.		

\subsection{Transformation of acceleration}

The gauge transformation of Sec.\ref{2C-lGt} defines the transformations:  
$\bx \to \bx'=\bx + \bxi$ and $\bv(\bx) \to \bv'(\bx')=\bv(\bx) +\DD_t\bxi$.  
Namely, the variations of  $x^i $ and  $v^i$ are given by 
\begin{eqnarray}
\del x^i & = & \xi^k T_k \,x^i = \xi^i, \hskip10mm (T_k = \pl/\pl x^k) \label{G-dq} \\
\del v^i  & = &  \DD_t \xi^i = \pl_t \xi^i  + v^k T_k\,\xi^i  \,,	\label{dGa} 
\end{eqnarray}
where $T_k x^i =\del_k^i$.  
Next, let us consider acceleration of a fluid particle.  Differentiating (\ref{lGtu})
with respect to $t$ again, we obtain 
\begin{equation}
\frac{\dd^2 }{\dd t^2}X'_a  = \frac{\dd^2 }{\dd t^2}X_a(t) + 
		\frac{\dd^2 }{\dd t^2}\bxi(X_a, t). 		\label{lGta}
\end{equation}
This describes a relation between accelerations in the two frames $F$ and $F'$. 
Let us define the covariant derivative of the velocity $v=(v^i)$ by 
$\DD_t v  =  \pl_t v  + v^k T_k\,v$. Variation of $\pl_t v$ is given by
\[ \del(\pl_t v) = \pl_t(\del v) + (\del \pl_t)v  		
   = \pl_t (\DD_t \xi) - (\pl_t \xi^k T_k) v  \,, 	
\]
where $\del \pl_t = \pl_{t'} - \pl_t =- \pl_t\xi^k \,T_k$ from (\ref{lGdt}) (to the 
first order of $\xi$).    Then, using $\del v= \DD_t \xi$ and $\del T_k = T'_k - T_k 
= - (T_k \xi^l)T_l$ from (\ref{lGdx}), the transformation of $\DD_t v$ is given by 
$\delta(\DD_t v) = \del(\pl_t v) + (\del v^k) T_k v + v^k (\del T_k) v
+ v^k T_k (\delta v)$, namely, 
\[ \delta(\DD_t v) = \pl_t (\DD_t \xi) - \pl_t \xi^k T_k v + (\DD_t \xi^k) T_k v 
	- v^k (T_k \xi^l)T_l v + v^k T_k (\DD_t \xi).    \]
Thus, we obtain
\[  \delta(\DD_t v) =  \DD_t(\DD_t \xi), \hskip20mm \DD_t=\pl_t +v^k T_k. 
\]
It is found that 
the covariant derivative $\DD_t v$ transforms just like the acceleration 
$\A=(\dd/\dd t)^2X_a(t)$ according to (\ref{lGta}).  This implies that $\DD_t v$ 
represents  the particle acceleration $\A(\bx,t)$:
\begin{equation}
\A= \DD_t \bv = (\pl_t  + v^k \pl_k) \bv .	\label{dtu-a} 
\end{equation}

\subsection{Galilean invariant Lagrangian }

According to the formulation of Kambe (2007) based on  the gauge principle, 
the total Lagrangian is defined by 
\begin{equation}
\Lambda_{\rm T}= \Lambda_{\rm f} + \Lambda_\eps +  \Lambda_\phi + \Lambda_\psi, 
\label{Lam-total} 
\end{equation}
as far as the translation symmetry is  concerned, where 
\begin{eqnarray}
\Lambda_{\rm f} & = & \int \half \langle \,\bv(x), \bv(x)\,\rangle \ \rho\, \dd^3\bx, 
 \hskip8mm \Lambda_\eps = - \int_{M} \, \eps(\rho, s) \ \rho \, \dd^3\bx , \label{GLdK} \\
\Lambda_\phi & = & - \int_{M}\, (\DD_t\phi)\,\rho \,\dd^3\bx, \hskip18mm    
\Lambda_\psi = - \int_{M}\, (\DD_t\psi)\,\rho\,s\, \dd^3\bx.   \label{L-g0}
\end{eqnarray}
The variables  $\bv$, $\rho$, $\eps$ and $s$ are the velocity vector, density, internal 
energy (per unit mass) and entropy (per unit mass) of the fluid, $\phi(\bx,t)$ and 
$\psi(\bx,t)$ are gauge potentials associated with the mass density $\rho$ and entropy 
density $\rho s$ respectively. The velocity is defined by $\bv(\bx,t)= \DD_t \bx$,
where $\DD_t =\pl_t + \bv\cdot \nabla$.  From \S\ref{2E}, it is almost 
obvious that the two Lagrangians $\Lambda_\phi$ and $\Lambda_\psi$ of (\ref{L-g0}) are 
invariant with respect to the local Galilean transformation (\ie local gauge 
transformation).

\section{Variational principle: \ Translation symmetry \label{4-VP1} }

\subsection{Action principle}

The action $I$ and the action  principle $\del I =0$ are defined by 
\begin{eqnarray}
I & = & \int_{t_0}^{t_1}  \Lambda_{\rm T}  \dd t = 
\int_{t_0}^{t_1}\int_M \dd t\,\dd^3\bx\ \,
		L_{\rm T}[\bv, \rho, s, \phi, \psi], \label{Action} \\
\del\,I & = & \del\,\int_{t_0}^{t_1}\int_M \dd t\,\dd^3\bx\ L_{\rm T}=0,\label{ActPri1} \\
&& L_{\rm T} \equiv \half\,\rho\,\langle\,\bv, \bv\,\rangle 
   - \rho\,\eps(\rho, s) - \rho\ \DD_t\phi - \rho s\ \DD_t\psi\,.  \label{L-total}
\end{eqnarray}
where $L_{\rm T}$ is the Lagrangian density.  There is  a certain thermodynamic 
property that must be taken into consideration.  That is the definition of an ideal 
fluid.  Namely, there is no dissipation of kinetic energy into heat, \ie there is no 
heat production within the ideal fluid. By thermodynamics, if there is no heat 
production, we have $T \del s=0$ ($T$: temperature). Then, 
\begin{equation}
\del \eps = (\del \eps)_s = \frac{p}{\rho^2}  \,\del \rho, \hskip5mm 
\big( \frac{\pl \eps}{\pl \rho}\big)_s = \frac{p}{\rho^2}, 
\hskip10mm \del h = \frac{1}{\rho} \del p, 		\label{del-e}
\end{equation}
where  $p$ is the fluid pressure, $h$ is the enthalpy defined by $\eps +p/\rho$,
and $(\,\cdot\,)_s$ denotes  $s$ being fixed.  However, the entropy $s$ may not be 
uniform and depend on $\bx$ through the initial condition. 

One of the aims of the present analysis is to show that the previous formulation 
(Kambe 2007) based on the symmetry of parallel translation alone is not sufficient 
to represent rotational flows with non-vanishing helicity. The rotaional 
symmetry will be considered from \S5 below.

\subsection{Outcomes of variations \label{3B}}

Writing the Lagrangian density  (\ref{L-total}) as $L_{\rm T}[\bv, 
\rho, s, \phi, \psi]$, we take variations of the  variables \ $\bv$, \ $\rho$, \ 
$s$ and potentials $\phi$ and $\psi$, where variations are assumed independent.
Substituting the varied variables \ $\bv+\del \bv$,\ $\rho +\del\rho$ \ $s +\del s$, 
\ $\phi +\del \phi$ and $\psi +\del \psi$ into $L_{\rm T}[\bv, \rho, s, \phi, \psi]$ 
and writing its variation as $\del L_{\rm T}$, we obtain 
\begin{eqnarray}
\del L_{\rm T} & = & \hskip4mm \delta \bv \cdot \rho\,(\,\bv - \,\nabla \phi
	- s\,\nabla \psi \,)  - \del s\ \rho\ \DD_t\psi   \nonumber \\
 	&&  + \ \delta \rho \ (\half v^2 - h - \DD_t \phi - s\,\DD_t \psi) \nonumber \\
	&&  + \ \del \phi \, \big(\, \pl_t \rho + \nabla\cdot (\rho \bv) \,\big) 
	    - \pl_t(\rho \,\del\phi) - \nabla \cdot(\rho\bv\,\del\phi)	\nonumber \\
	&&  + \ \del \psi \, \big(\, \pl_t (\rho s)+ \nabla\cdot (\rho s \bv) \,\big) 
	    - \pl_t(\rho s\,\del\psi) - \nabla \cdot(\rho s\bv\,\del\psi), \label{d-Ld}
\end{eqnarray}
where $(\pl/\pl \rho)(\rho\,\eps) = \eps+\rho\,(\pl \eps/\pl \rho)_s = \eps +p/\rho=h$
is used. The variation  fields are assumed  to vanish on the boundary surface enclosing 
the domain $M$, as well as at both ends  $t_0$ and $t_1$ of time integration 
of the action  $I$ (where $M$ is chosen arbitrarily), as is usually the case.
Thus, the action principle $\delta I =0$ for independent variations 
$\delta \bv$, $\delta \rho$ and $\delta s$ results in
\begin{eqnarray}
\delta \bv &:&  \hskip5mm   \bv = \nabla\, \phi + s\, \nabla \psi\,, \label{dvel}  \\
\delta \rho &:& \hskip5mm \half v^2 -h - \DD_t \phi - s\,\DD_t \psi=0\,, \label{drho1} \\
  \del s &:& \hskip5mm \DD_t\psi \equiv \pl_t\psi + \bv\cdot \nabla \psi =0\,. 
\label{del-s} 	
\end{eqnarray}
Using (\ref{dvel})  and (\ref{del-s}),  we have 
\[	\DD_t\phi= \pl_t\phi + \bv\cdot \nabla \phi = \pl_t\phi 
	+ \bv\cdot (\bv - s \nabla \psi) = v^2 + \pl_t\phi + s \,\pl_t\psi. \]
Under this relation and (\ref{del-s}), the equation (\ref{drho1}) can be rewritten as
\begin{equation} 
\half v^2  + h + \pl_t \phi +  s\,\pl_t \psi=0 \,. \label{drho2}
\end{equation}
This is regarded as an integral of motion.  From the variations of $\delta \phi$ 
and $\delta \psi$, 
\begin{eqnarray}
  \del \phi   &:&  \hskip5mm \pl_t \rho + \nabla\cdot(\rho \bv) =0\,, \label{dphi} 
\\[1mm]
  \del \psi   &:&  \hskip5mm \pl_t(\rho s)+ \nabla\cdot(\rho s \bv) =0\,. \label{dpsi1}
\end{eqnarray}
Using (\ref{dphi}), the second equation  reduces to the {\it adiabatic} equation:
\begin{equation} 
\pl_t s + \bv \cdot \nabla s = \DD_t s = 0\,. \label{dpsi2}
\end{equation}
Thus, we have obtained the  continuity equation 
(\ref{dphi}) and  entropy equation (\ref{dpsi2}) from the action principle.
 
It is shown in Kambe (2007) that the present solution is equivalent to the classical 
Clebsch solution (Clebsch 1859).  With the velocity (\ref{dvel}), 
the vorticity $\bom$ is given by 
\begin{equation} 
\bom = \nabla \times \bv = \nabla s \times \nabla \psi.	 \label{om-sp}
\end{equation}	
This implies that the vorticity is connected with non-uniformity of entropy.  Using  
(\ref{drho2}), we find that the following Euler's equation of motion is satisfied:
\begin{equation} 
\pl_t \bv + \bom \times \bv = -\nabla \big(\half v^2+ h \, \big).  \label{Euler-eq1} 
\end{equation}	
In this case, the helicity vanishes (Bretherton, 1970). 
For a general velocity field, the helicity $H$ defined by 
\begin{equation} 
H[V]  \equiv  \int_V \ \bom \cdot \bv \ \dd^3\bx    \label{Hel-V}
\end{equation}  
is a measure of linkage and knottedness of  vortex lines, and does not vanish 
in general (see Sec.\ref{8A}).

\subsection{Homentropic fluid \label{3C}}

For a homentropic fluid in which the entropy $s$ is a uniform constant $s_0$ at all 
points, we have $\eps=\eps(\rho)$, \ $\dd \eps= (p/\rho^2)\,\dd \rho$ and $\dd h 
= (1/\rho) \dd p$  from (\ref{del-e})  since $\del s=0$.  In addition, 
the motion is {\it irrotational}. In fact, from (\ref{dvel}), we have $\bv = 
\nabla\, \Phi$ where $\Phi=\phi + s_0\, \psi$,  and $\bom=0$ from (\ref{om-sp}). 
The integral (\ref{drho2}) becomes $\half v^2  + h + \pl_t \Phi =0$.  The  Euler's 
equation (\ref{Euler-eq1}) reduces to 
\begin{equation}
\pl_t \bv + \nabla ( \half v^2) = - \nabla \, h \,,  \hskip10mm \mbox{where}\hskip5mm
	\nabla \,h  = \frac{1}{\rho}\, \nabla \,p \,.	  \label{Euler1} 
\end{equation}
Note that the left hand side   is the convective time derivative for the irrotational  
velocity $v^k=\pl_k\Phi$:
\begin{equation}
\pl_t \bv + \nabla(\half v^2) =\pl_t \bv+(\bv\cdot\nabla)\bv= \DD_t\bv \,.\label{MDpot} 
\end{equation}
Thus, as far as the action principle is concerned for a homentropic fluid, Euler's 
equation of motion reduces to that for {\it potential} flows of a perfect fluid.  

Traditionally,  this property is considered to be a defect of the formulation 
of the Eulerian variation described in the previous section, because the action 
principle should yield the equations for rotational flows as well.  In order to 
remove this (apparent) flaw,  Lin (1963) introduced the condition for the 
conservation of the identity of particles denoted by $\ba=(a^k)$, which is represented 
by an additional subsidiary Lagrangian of the form $\int\,A_k \cdot\DD_t a^k\ \dd^3\bx$. 
This introduces three potentials $A_k(\bx,t)$ as a set of Lagrange multipliers of 
conditional variation.  Problem is that physical significance of $A_k$ is not clear 
(Bretherton 1970;  Seliger \& Whitham 1968; Salmon 1988).

Let us recall that we have been considering the Lagrangian $\Lambda_{\rm T}$ satisfying 
the symmetry of parallel translation. However, the flow field has another symmetry of 
rotational invariance (Kambe 2003a, b).  The equation (\ref{dpsi2}) implies that 
the entropy $s$ plays the role to identify each fluid particle.  Furthermore, local 
rotation is captured by the expression (\ref{om-sp}).   However, in a homentropic 
fluid, there is no such machinery to identify each fluid particle. Gauge invariance 
with respect to local rotation could be a candidate instead of $s$. 

Insofar as the flow field is characterized by the translation symmetry alone, 
we have arrived at the above result, \ie the flow field should be {\it irrotational} 
if the fluid is homentropic.  The fluid motion is driven by the velocity potential 
$\Phi$, where $\Phi = \phi + s_0\,\psi$ with $\phi$ and $\psi$  the gauge potentials.

\section{Equations in  $\ba$-space \label{4-FLV} }

\subsection{Lagrangian \label{4A}}

Let us consider another variational formulation with the Lagrangian represented by 
the particle coordinates $\ba=(a^1,a^2,a^3)=(a,b,c)$. Independent variables are 
denoted with $a^\mu$ ($\mu=0,1,2,3$) with $a^0$ being the time variable written 
as $\tau$ ($=t$). The letter $\tau$ is used instead of $t$ in combination 
with the particle coordinates $a^k$ ($k=1,2,3$). The physical-space position of 
a particle is expressed by $X^k(a^\mu)$, or $(X,Y,Z)$. Its velocity $v^k$ is 
given by $\pl_\tau X^k = X_\tau^k$. The derivative $\pl_\tau=\pl/\pl \tau$ 
is equivalent to the covariant derivative $\DD_t$ :
\begin{equation}
\pl_\tau = \pl_t + u\pl_x + v\pl_y +w\pl_z = \pl_t + \bv\cdot\nabla \,. \label{dt-Dt}
\end{equation}
Mass density $\rho$ is defined by the relation, $\rho\,\dd^3\bx=\dd^3\ba$, 
where $\dd^3\bx$ denotes the volume element $\dd X^1\dd X^2\dd X^3$.  Using the Jacobian 
determinant $J$ of the transformation from $\ba$-space to $\bX$-space, we obtain 
\begin{equation}
\rho = \frac{1}{J}, \hskip10mm J = \frac{\pl(X^1,X^2,X^3)}{\pl(a^1,a^2,a^3)}
	= \frac{\pl(X, Y, Z)}{\pl(a, b, c)}\,.                     \label{Jacob}
\end{equation}
The total Lagrangian (\ref{Lam-total}) with (\ref{GLdK}) and (\ref{L-g0})
is rewritten as 
\begin{eqnarray}
\Lambda_{\rm T}  & = & \int \half\, X_\tau^k\,X_\tau^k\, \dd^3\ba - \int \eps(\rho,s)\,
 \dd^3\ba - \int \pl_\tau\phi\,\dd^3\ba - \int s\, \pl_\tau\psi\,\dd^3\ba. \label{LT-2}
\end{eqnarray} 
The action $I$ is defined by integration of $\Lambda_{\rm T}$ with 
respect to the time $\tau$: 
\[   I = \int_{\tau_1}^{\tau_2} \Lambda_{\rm T} \dd\tau.  \]
In this definition, the third integral $I_3=\int \dd\tau \int \pl_\tau\phi \,\dd^3\ba$ 
can be integrated with respect to $\tau$, and expressed as $\int [\phi] \dd^3\ba$, where 
$ [\phi]= \phi|_{\tau_2} - \phi|_{\tau_1}$ is the difference of $\phi$ at the ends 
$\tau_2$ and $\tau_1$ and is independent of $\tau \in [\tau_1,\tau_2]$.  Likewise, the 
fourth integral can be expressed as $I_4=\int [\psi] s\, \dd^3\ba$, 
because the entropy $s$ is  independent of $\tau$ according to (\ref{dpsi2}). This means 
that the gauge potentials $\phi$ and $\psi$ do not appear in the variations of the action 
$I$  for $\tau \in [\tau_1,\tau_2]$.  In other words, the gauge potentials $\phi$ and 
$\psi$ make their appearance only in the action 
represented with the physical space 
coordinates $\bx$.  The fact that the last two Lagrangians in (\ref{LT-2}) do not 
contribute to the variations (for $\tau \in [\tau_1,\tau_2]$) implies that local 
Galilean symmetry is in fact an internal symmetry.

Thus, by omitting the integrated terms, we have 
\begin{eqnarray}
I  & = & \int\, L(X_\mu^k) \,\dd^4 a,	\hskip10mm 
	L = \half\,X_0^k\,X_0^k - \eps(X^k_l,a^k) , \label{I-LdT}
\end{eqnarray}
where $X^k_l=\pl X^k/\pl a^l$, \ $X_0^k = X_\tau^k =v^k$ \ ($k, l=1,2,3$), and 
$\dd^4 a= \dd \tau\,\dd^3\ba $.  In the $\ba$-space, the point $\ba$ is motion-less, 
while in the $\bx$-space,  there is a flow of particles driven by the potentials 
$\phi$ and $\psi$ according to (\ref{dvel}).

\subsection{Noether's conserved currents \label{4B}}

The Euler-Lagrange equation can be writtten as
\begin{equation}
\frac{\pl}{\pl a^\mu} \Big( \frac{\pl L}{\pl X^k_\mu} \Big) - \frac{\pl L}{\pl X^k} 
= \pl_\mu \Big(\frac{\pl L}{\pl X^k_\mu} \Big) -\frac{\pl L}{\pl X^k} =0 \label{EL-X}
\end{equation} 
($\pl_\mu=\pl/\pl a^\mu$).  The energy-momentum tensor $T_\mu^\nu$ is defined by 
\[ T_\mu^\nu \equiv X_\mu^k\, \Big( \frac{\pl L}{\pl X_\nu^k} \Big) 
		- L\,\del_\mu^\nu \,,  \]	
where $k=1,2,3$. As long as (\ref{EL-X}) is valid together with $\pl_\tau L=0$, 
we have a conservation equation $\pl_\nu T_\mu^\nu=0$. In fact, we have
\begin{eqnarray*}
\pl_\nu T_\mu^\nu  & = & \pl_\nu(X_\mu^k) \Big( \frac{\pl L}{\pl X_\nu^k} \Big) 
	+ X_\mu^k \,\pl_\nu \Big( \frac{\pl L}{\pl X_\nu^k} \Big) - \pl_\mu X^k 
	\frac{\pl L}{\pl X^k} - \pl_\mu (X_\nu^k) \frac{\pl L}{\pl X_\nu^k}    \\
	& = & X_\mu^k \, \left[ \pl_\nu \big( \frac{\pl L}{\pl X_\nu^k} \big) 
		- \frac{\pl L}{\pl X^k}\right] = 0
\end{eqnarray*}
This is the Noether theorem (Noether (1918), Weinberg (1995)).

By using the Lagrangian $L$ of (\ref{I-LdT}), we obtain $\pl L/\pl X_0^k =X_0^k 
=v^k$.  The internal energy $\eps$ depends on the density $\rho$ which in turn 
depends on $X^k_l=\pl X^k/\pl a^l$ by (\ref{Jacob}), where the entropy $s$ depends on 
$a^l$ only.  Hence, we have 
\begin{equation} 
\frac{\pl L}{\pl X_l^k} = -\frac{\pl \eps}{\pl \rho} \frac{\pl \rho}{\pl X_l^k}
	= - \frac{p}{\rho^2} \frac{\pl \rho}{\pl X_l^k}  \,,   \label{dLdX}
\end{equation}
by using the thermodynamic relation given  in (\ref{del-e}).  In view of 
(\ref{Jacob}), we obtain 
\begin{equation}
\frac{\pl L}{\pl X_1^1} = p \, \frac{\pl (X^2,X^3)}{\pl (a^2,a^3)}, \hskip5mm
\frac{\pl L}{\pl X_1^2} = p \, \frac{\pl (X^3,X^1)}{\pl (a^2,a^3)}, \hskip5mm
\frac{\pl L}{\pl X_1^3} = p \, \frac{\pl (X^1,X^2)}{\pl (a^2,a^3)}. \label{dLdXkl}
\end{equation}
Thus, the energy-momentum tensor $T_\mu^\nu$ are found as follows (Eckart 1960):
\begin{eqnarray} 
T_0^0  & = & X_\tau^k\, \Big( \frac{\pl L}{\pl X_\tau^k} \Big) - L = X_\tau^k X_\tau^k 
   - L = \half\,v^2 + \eps \equiv H \hskip2mm \mbox{(energy density)}, \label{T0mu} \\
T_0^1 & = & X_\tau^k\, \Big( \frac{\pl L}{\pl X_1^k} \Big) = p \, 
	\frac{\pl (X, Y, Z)}{\pl (\tau,b, c)},  \hskip20mm 
	(\pl\eps/\pl\rho = p/\rho^2), 	\nonumber   \\
T_0^2 & = & p \, \frac{\pl (X, Y, Z)}{\pl (a,\tau,c)},  \hskip20mm
T_0^3  =  p \, \frac{\pl (X, Y, Z)}{\pl (a,b,\tau)} \nonumber
\end{eqnarray}
Introducing the symbols $(\al,\beta,\gamma)$ to denote $(a,b,c)$ cyclically, another 
twelve components are
\begin{eqnarray} 
T_{\al}^0 & = & X_{\al}^k\, \Big( \frac{\pl L}{\pl X_0^k} \Big) = X_{\al}^k\,X_\tau^k 
	= X_{\al} X_\tau + Y_{\al} Y_\tau + Z_{\al} Z_\tau,\ \  \equiv V_{\al},  
	 \label{Ta0}    \\
T_{\al}^{\al} & = & X_{\al}^k\, \Big( \frac{\pl L}{\pl X_{\al}^k} \Big) - L 
	= p\, \frac{\pl (X, Y, Z)}{\pl (\al,\beta, \gamma)} - (\half\,v^2 - \eps)
	= - \half\,v^2 + h,    \label{Taa}    \\
T_{\al}^{\beta} & = & X_{\al}^k\, \Big( \frac{\pl L}{\pl X_{\beta}^k} \Big) = 0, 
	\hskip20mm 	T_{\al}^{\gamma}  =  0.  \nonumber
\end{eqnarray}
The Noether conservation law $\pl_\nu T_\mu^\nu=0$ reduces to the momentum equations 
(see below) for $\mu=\al$: 
\begin{equation}
\pl_\tau V_{\al} + \pl_{\al} \,F = 0  \hskip10mm (\mbox{for $\al=a, b,c$}), 
\label{EqM-a-1}
\end{equation}
where $F=- \half\,v^2 + h$, and other two equations are obtained with cyclic permutation 
of $(a,b,c)$. Integrating this with respect to $\tau$ between the limits 0 and $t$, 
we find the Weber's transformation (Lamb 1932, Art.15):
\begin{equation}
V_{\al}(\tau) =  X_{\al} X_\tau + Y_{\al} Y_\tau + Z_{\al} Z_\tau  
= V_{\al}(0) - \pl_{\al} \chi,  \label{Web-Tr}	
\end{equation}
where  $\chi = \int_0^t \, F\,\dd\tau = \int_0^t (- \half\,v^2 + h) \dd\tau$.
For $\mu=0$, the conservation law $\pl_\nu T_\mu^\nu=0$ describes the energy equation:
\begin{equation}
\pl_\tau H + \pl_a \left[ p \frac{\pl (X, Y, Z)}{\pl (\tau, b, c)} \right]
	   + \pl_b \left[ p \frac{\pl (X, Y, Z)}{\pl (a, \tau, c)} \right]
	   + \pl_c \left[ p \frac{\pl (X, Y, Z)}{\pl (a, b, \tau)} \right] = 0.  
\label{CEq-0}
\end{equation}
In the Eulerian description with the independent variables $(t,x,y,z)$ instead of 
$(\tau,a,b,c)$, this reduces to 
\[   \DD_t (\half \,v^2 +h )  =  \frac{1}{\rho}\,\pl_t p.  \]
This can be transformed to the following form of conservation of energy:
\begin{equation}
\pl_t (\half \,v^2 + \eps )  +\pl_k \left[ (\half \,v^2 +h)\,v^k \right]=0. \label{CEq-0-x}
\end{equation}
Using (\ref{Ta0}) and (\ref{Taa}),  the equation (\ref{EqM-a-1}) reduces to the 
equation for the acceleration $A_{\al}(\tau,\ba)$:
\begin{equation}
A_{\al} \equiv X_{\al} X_{\tau\tau} + Y_{\al} Y_{\tau\tau} + Z_{\al} Z_{\tau\tau} 
	= - \frac{1}{\rho}\,\pl_{\al} p ,  \label{EqM-a-2}
\end{equation}
which is known as the Lagrangian form of the equation of motion (Lamb 1932, Art.13).
This is obtained by noting that $\pl_{\al} h=(1/\rho)\pl_{\al} p$  and 
\[ X_{\al\tau} X_\tau + Y_{\al\tau}Y_\tau + Z_{\al\tau}Z_\tau = \pl_{\al}(\half v^2). \]
In view of (\ref{dtu-a}),  the equation (\ref{EqM-a-2}) is equivalent to  
Euler's equation of motion in the Eulerian description of the barotropic relation 
$h(p)=\int^p \dd p'/\rho(p')$:
\begin{equation}
\pl_t \bv + (\bv\cdot\nabla) \bv = - \frac{1}{\rho}\,\nabla\,p = -\nabla\,h. \label{EEq}
\end{equation} 

Velocities $V_{\al}(\tau,\ba)$  in the $\ba$-space may be determined by (\ref{Web-Tr}) 
with given initial conditions of $V_{\al}(0,\ba)$ and $h(0,\ba)$ at $\ba=(a,b,c)=\bx$. 
However, in order to transform it to the $\bx$-space, there is some freedom.
The equation (\ref{Web-Tr}) is invariant by rotational transformations 
of a displacement vector $\Del \bX=(\Del X, \Del Y, \Del Z)$ since the left hand side is
of the form of a scalar product with respect to $\Del \bX$, so that the vector $\Del\bX$ 
is not uniquely determined (the density $\rho$ and enthalpy $h$ are not changed by this 
transformation since it is volume-preserving). The same freedom applies to the 
acceleration $A_{\al}(\tau,\ba)$ of (\ref{EqM-a-2}) as well.  Rotational transformations 
are considered in the next section. This formulation helps to eliminate the arbitrariness.

\section{Rotational transformations \label{5-RT}}

Next, in order to equip the action $I$ with an additional mathematical structure of 
rotational symmetry (to be considered in \S\ref{S6}), we summarize the rotational 
transformations here, and consider the rotational gauge symmetry. The related 
gauge group is the rotation group $SO(3)$.  An infinitesimal rotation is 
described by the Lie algebra  ${\bf so}(3)$ of three dimensions. The basis 
vectors of  ${\bf so}(3)$ are denoted by $(e_1, e_2, e_3)$, which satisfy the 
commutation relations:
\begin{equation}
	[\, e_j, \, e_k\,] = \vep_{jkl}\, e_l  \,,  	\label{Rcomm}
\end{equation}
where $\vep_{jkl}$ is the completely skew-symmetric third-order tensor, and 
\begin{equation} 
e_1 =\left[ \begin{array}{ccc}  0 & 0 & 0 \\ 
	0 & 0  & - 1  \\  0  &  1 & 0 \end{array} \right],\hskip5mm
e_2 =\left[ \begin{array}{ccc}  0 & 0  & 1  \\ 
	0 & 0  & 0  \\  - 1  &  0  & 0 \end{array} \right],\hskip5mm
e_3 =\left[ \begin{array}{ccc}  0 & - 1 &  0  \\ 
	1 & 0  &  0  \\   0  &   0  & 0 \end{array} \right].	\label{RT2}
\end{equation}
A rotation operator is defined by $\theta = \theta^k e_k$ where $\theta^k$
($k=1,2,3$) are infinitesimal parameters: 
\[  \theta  =(\theta_{ij}) \ \equiv\ \theta^k\,e_k = \left[ 
	\begin{array}{ccc}  0 & -\theta^3 & \theta^2 \\ 
  \theta^3 & 0  & - \theta^1  \\  -\theta^2  &  \theta^1 & 0 \end{array} \right]\,. \]
Then, an infinitesimal rotation of the displacement vector $\bs=(s^1, s^2, s^3)$ is 
expressed by $ \theta\,s$. This is also written as  $\hat{\theta} \times \bs$, where 
$\hat{\theta} = (\theta^1,\,\theta^2,\,\theta^3)$ is an infinitesimal angle vector.  

Preliminary studies of the rotational symmetry of fluid flows are given in 
Kambe (2003a, b).  However, more details are elaborated in the following 
subsections \ref{5A} $\sim$ \ref{5C}, and the section \ref{S6} proposes
a new Lagrangian $L_A$ to account for the rotation symmetry.

\subsection{Gauge transformation (Rotation symmetry) \label{5A}}

We consider local rotation of a fluid element about an arbitrary reference point $\bx_0$ 
within a fluid.  The non-abelian property of the rotational transformation  requires
infinitesimal analysis.  At a neighboring point  $\bx_0+\bs$ for a small $\bs$, 
we consider its local transformation, $\bs \to \bs+\del\bs$.

Rotational gauge transformation of a point $\bs$ in a frame $F$ (an inertial frame, say)
to $\bs'$ in a non-inertial frame $F'$ is expressed as 
\begin{equation}
 \bs'(\bs,t) = \bs + \del \bs(\bs,t) , \hskip10mm t' = t,    \label{lRts}	
\end{equation}
instead of (\ref{lGtx}) for the translation. The variation $\del \bs$ is defined by an 
infinitesimal rotation  $\theta= \theta(t)$ (skew-symmetric) applied to the 
displacement vector $\bs$: $\del \bs \equiv \bta(\bs,t) = \theta \bs 
= \hat{\theta} \times \bs$.  The corresponding transformation of velocity 
($\bv=\DD_t\bs$) is obtained from (\ref{lRts}) as follows: 
\begin{eqnarray}
\bv' (\bs') =\DD_t\bs' & = & \DD_t(\bs + \del \bs) = \bv(\bs) + \DD_t \bta, \label{lRtv} \\
\del \bv(\bs)  & = &  \DD_t \bta \ \ \equiv \ \pl_t\bta(\bs, t) + (\bv\cdot \nabla_s 
	\big) \bta(\bs,t) = \pl_t\theta\,\bs + \theta \bv,  \label{lRxi}
\end{eqnarray}
instead of (\ref{lGtu}) and (\ref{lGxi}) where $\nabla_s=(\pl/\pl s^i)$.  As before 
(Sec.~\ref{2C-lGt}), in this transformation  the points $\bs$ 
and $\bs'$ are the same point with respect to the frame $F$. This means that
the  coordinate frame $F'$ is rotated by an angle $-\hat{\theta}$ with respect to $F$
with a fixed origin $\bx_0$. At the origin $\bx_0$ ($\bs=0$), we have the following:
\begin{equation} 
\del \bv(0) = \theta \,v = \theta^k e_k v ,   \label{dv0} 	
\end{equation}
from (\ref{lRtv}) and (\ref{lRxi}), where $v =\bv(\bx_0)$.

\subsection{Acceleration}

Let us consider the time-derivative  $\pl_t \bv$ in the Euler equation 
(\ref{Euler1}) for a homentropic fluid.  It is proposed that this term be 
replaced with $\dd_t v$ as 
\begin{equation} 
\pl_t v \ \to \   \dd_t v = \pl_t v + \Om \,v, \hskip15mm 
\Om \,v = \Om^k e_k \,v = \hat{\Om}  \times v,  \label{Om-v} 	
\end{equation}
where $\Om$ is an operator defined by $\Om = \Om^k e_k$ for scalar fields $\Om^k$, 
and $\hat{\Om}$ is an axial vector defined by $(\Om^1,\,\Om^2,$ $\Om^3)$.\footnote{A 
simple font $v$ is used here for brevity instead of bold-face three-vector $\bv$.}
The tensor $\Om_{ij}\equiv (\Om^k e_k)_{ij}$  is skew-symmetric  ($\Om_{ij}=- \Om_{ji}$).

We consider variations  at the origin ($\bs=0$).  Then, we have $\del v= \theta^k e_k v$ 
from (\ref{dv0}).  Variation of $\Om^k$ is defined by\footnote{The vector form is \ 
$\del \hat{\Om} = \pl_t\hat{\theta} + \hat{\theta} \times \hat{\Om} $.} 
\begin{equation}
\delta \Om^k = \pl_t \theta^k + \vep_{klm} \theta^l\,\Om^m \,. \label{dOm}    
\end{equation}
Using $\del \pl_t = \pl_{t'} - \pl_t = - \pl_t\eta^k \,T_k$ from (\ref{lGdt}),  
we obtain  the variation of $\dd_t v$:
\[  \delta(\dd_t v) = \pl_t(\del v) + (\del \pl_t) v 
	+ \del \Om^k\,e_k\,v  + \Om^k e_k \,\del v
	= \pl_t(\theta^k e_k \,v) - \pl_t (\theta\,s)^kT_k v   \]   
\[  + (\pl_t \theta^k + \vep_{klm} \theta^l \,\Om^m ) e_k\,v 
	+ \Om^m e_m \, \theta^l e_l \,v ,	\]    
where  $\vep_{klm}$ can be eliminated by using the following equality obtained from 
(\ref{Rcomm}): $\vep_{lmk} e_k$ $= e_l e_m - e_m e_l$. Setting $s=0$, we obtain
\begin{equation}
\delta(\dd_t v) = \theta^k e_k\,[\pl_t v+ \Om^m e_m \,v ] + 2\pl_t \theta^k\,e_k\,v 
	= \hat{\theta} \times \dd_t v + 2\,\pl_t\hat{\theta} \times v.  \label{dNtv2}
\end{equation}
The second term of the last expression denotes the Coriolis term 
in the rotating system.  Thus, it is found that the covariant derivative 
$\dd_t v$ behaves like the acceleration in a rotating frame, as far as $\Om$ 
transforms according to  (\ref{dOm}).

In the previous analysis for the translation symmetry, it was shown 
that the left hand side of (\ref{Euler1}), $\pl_t \bv + \nabla ( \half v^2)$, 
is in fact equal to the particle acceleration $\DD_t\bv$ of (\ref{MDpot}). 
In the present section taking into account the rotation symmetry, the partial 
derivative $\pl_t$ is replaced with the covariant derivative $\dd_t=\pl_t + \Om$.  
Thus, the particle acceleration  $\nabla_t\bv$ should be defined by
\begin{equation}   
\nabla_t \,\bv \equiv \pl_t \bv + \Om \,\bv +\grad( v^2/2) 
	= \pl_t \bv + \hat{\Om}  \times \bv + \grad( v^2/2)  \,, \label{CDrot}
\end{equation}
where $\hat{\Om}$ is the gauge field with respect to the rotation symmetry.

\subsection{Vorticity as a gauge field \label{5C}} 

It can be verified that the gauge field $\hat{\Omega}$ in fact coincides with the 
vorticity $\nabla \times \bv$ by the requirement of Galilean invariance of the 
covariant derivative $\nabla_t \bv$ of (\ref{CDrot}), as follows.
Under a global Galilean transformation from one frame $F:\ x = (t,\ \bx)$ to another 
$F_*:\ x_*= (t_*,\ \bx_*)$ which is moving with  a uniform velocity $\bU$ relative to 
$F$, the position vector $\bx_*$  and velocity vector $\bv_*$ in the 
frame $F_*$ are given by $\bx_*= \bx - \bU t$ and $\bv_* =\bv - \bU$.
Transformation laws of derivatives  are $\pl_t = \pl_{t_*} -\bU\cdot \bnabla_*$ and   
$\bnabla = \bnabla_*$.  Applying these and  replacing $\bv$  by $\bv_*+\bU$,  
the covariant derivative of (\ref{CDrot}) is transformed to 
\begin{eqnarray*}	
   && (\pl_{t_*} - \bU\cdot \nabla_*) (\bv_*+\bU) + \hat{\Omega} \times (\bv_*+\bU)
		+ \nabla_*\,\half \,|\bv_*+\bU|^2	   \\
 & = &  \ \pl_{t_*} \bv_* + \hat{\Omega} \times \bv_* + \nabla_*\, (v_*^2/2)		
 	-\, (\bU\cdot \nabla_*)\bv_* + \hat{\Omega} \times \bU 
	+ \nabla_*\big( \bv_*\cdot \bU \big) \,,	
\end{eqnarray*}   
since $\bU$ is a constant vector and $\nabla_*(U^2)=0$.  We require the covariance, \ie 
$\nabla_t \bv = (\nabla_t \bv)_*$. Namely, the right hand side should be equal to 
$\pl_{t_*} \bv_* + \hat{\Omega}_* \times \bv_* + \nabla_* (v_*^2/2)$.  Therefore,  
\begin{eqnarray}	
0 & = & (\hat{\Omega}-\hat{\Omega}_*) \times \bv_* - (\bU\cdot \nabla_*)\bv_* 
+ \hat{\Omega} \times \bU 
	+ \nabla_*\big( \bv_*\cdot \bU \big) 	\nonumber   \\
& = & (\hat{\Omega}-\hat{\Omega}_*) \times \bv_*  + ( \hat{\Omega}- 
\nabla_* \times \bv_*) \times \bU, 
\label{GFGI}
\end{eqnarray}
on using the following vector identity: $\bU \times (\nabla_*\times\bv_*) 
= -(\bU\cdot \nabla_*)\bv_* + \nabla_*(\bU\cdot\bv_*)$, for the constant vector $\bU$.  
Equation (\ref{GFGI}) is satisfied identically, if
\begin{equation} 
\hat{\Omega} = \hat{\Omega}_*\,, \hskip15mm \hat{\Omega} = \nabla \times \bv
		= \nabla_* \times \bv_* =  \hat{\Omega}_*   \label{OmVor}  \,.	
\end{equation}
The second relation holds by the Galilean transformation since $\nabla=\nabla_*$ and 
$\bv=\bv_*+\bU$.   Thus, the Galilean invariance of $\nabla_t \bv$ results in 
\begin{equation} 
 \hat{\Omega} = \nabla \times \bv \equiv \bom \,.   \label{OmVor2}  
\end{equation}
Namely, the  gauge field $\hat{\Omega}$ {\it coincides} with the  vorticity $\bom$.
Consequently,  the covariant derivative $\nabla_t \bv$ (particle acceleration) is 
given by 
\begin{equation}
\nabla_t \bv = \pl_t \bv + \nabla(\half\, v^2) + \bom \times \bv .	\label{CDrot*}
\end{equation}
Using the  vector identity
\begin{equation}
\bv \times (\nabla \times \bv) = \nabla (\half |\bv|^2) - (\bv \cdot \nabla)\bv, 
\label{VI-S63}
\end{equation}
the last expression (\ref{CDrot*}) is transformed to the convective derivative of $\bv$ 
(\ie the particle acceleration):
\begin{equation}
\nabla_t \bv = \pl_t \bv + (\bv \cdot \nabla)\,\bv .	\label{CDrot*1}
\end{equation}
Thus, we have $\nabla_t  = \pl_t  + (\bv \cdot \nabla)$.
So far, $\bx_0+\bs$ was a local position vector around a fixed point $\bx_0$.  
In general,  using $\bx$ in place of $\bx_0+\bs$, we arrive at the same definition:
$\bv(\bx,t) = \nabla_t \bx =  (\pl_t  + (\bv \cdot \nabla))\bx$.

\section{Lagrangian $L_A$ associated with rotation symmetry \label{S6}}

Associated with  the rotation symmetry, we try to introduce an additional Lagrangian
according to the  gauge principle.  The property of $I_3$ and $I_4$ considered 
below the equation (\ref{LT-2})  suggests how to find 
such a Lagrangian.  It was seen there that the Lagrangians represented with the 
particle coordinates $\ba=(a^1,a^2,a^3)=(a,b,c)$ can be integrated with respect to 
the time $\tau$ and  eliminated in the action to deduce the Euler-Lagrange equation, 
while they are non-trivial in the physical-space coordinates $\bx=(x^1,x^2,x^3)=(x,y,z)$ 
because of the non-trivial Jacobian matrix $(\pl x^k/\pl a^l)$ of the transformation.  
As a result, the action principle yields the continuity equation and entropy equation.
Thus the gauge potentials are regarded as {\it mathematical} agents with mechanical 
functions that drive fluid motions in the physical $\bx$-space, whereas 
they disappear in motionless  $\ba$-space.  

It is important to observe that, in the Lagrangian (\ref{LT-2}), 
the integrands of the last two integarals are of the form $\pl_\tau(\,\cdot\,)$, 
because the entropy $s$ and mass element $\rho \dd^3\bx=\dd^3\ba$ are independent 
of $\tau$.  The action is defined by $I=\int\int [L'+ \pl_\tau(\,\cdot\,)]\,
\dd\tau \dd^3\ba$, where $L'$ denotes the part of the first two terms of (\ref{LT-2}). 
This property is the simplest case of representation of topology in the gauge theory.
In the context of rotational flows, it is known that the helicity (or Hopf invariant,
Arnold \& Khesin (1998)) defined by (\ref{Hel-V}) describes non-trivial
topology of vorticity field, \ie mutual linking of vorticity lines.  This 
is closely related with the Chern-Simons term in the gauge theory. The Chern-Simons term 
lives in one dimension lower than the original four-space-time ($x^\mu$) of the action 
$I$ because a topological term in the action is expressed in a form of total divergence 
($\pl_\mu F^\mu$) and characterizes topologically non-trivial structures of the gauge 
field (Chern 1979; Jackiw 1985; Desser {\it et al.} 1982). 
Here, we follow  the formulation of Sec.\ref{4A} and look for a 
$\tau$-independent field directly.

\subsection{Lagrangian in  $\ba$-space   \label{6A}}

The $\tau$-independent field can be found immediately from Eq.~(\ref{EqM-a-1}).  
Taking the curl of this equation with respect to the coordinates $(a,b,c)$, we  obtain 
\begin{equation}
\nabla_a \times \pl_\tau \bV_a = \pl_\tau (\nabla_a \times \bV_a)= 0 , \label{dtOm-a}
\end{equation}
since $ \nabla_a \times \nabla_a F=0$, where $\nabla_a=(\pl_a, \pl_b, \pl_c)$, and 
$\bV_a=(V_a, V_b,V_c)$ is defined by (\ref{Ta0}). It is found that 
$\nabla_a \times \bV_a$ is  independent of $\tau$.\footnote{This property can be 
related to the invariance of the Lagrangian with respect to the relabeling
transformation $\ba \to \ba'$, satisfying the mass invariance $\dd^3 \ba'
=\dd^3\ba$.}  Hence, one may write as
$\nabla_a\times \bV_a= \bOm_a(\ba)$, where the right hand side is a vector 
depending on $\ba$ only.  It is useful to formulate this property in the framework
of exterior differential forms and Lie derivative.
Vector $\bV_a$ is understood as a transformed form of the velocity 
$\bv=(X_\tau,Y_\tau,Z_\tau)=(u,v,w)$  into the $\ba$-space.  This is seen on the 
basis of a 1-form $\V$ defined by
\begin{eqnarray}
\V      & = & V_a \,\dd a + V_b \,\dd b + V_c \,\dd c  \hskip5mm 
	(\mbox{written as} \ \ \bV_a \cdot \dd \ba)  \label{V1aa} \\
        & = & u\,\dd x + v \, \dd y + w \,\dd z. \hskip5mm \label{V1xx} 
\end{eqnarray}  
where $V_a= u x_a + v y_a + w z_a$, $x_a=\pl X/\pl a$, $u=X_\tau$, \etc and 
$\dd \ba=(\dd a, \dd b, \dd c)$. Hence $V_a$ is equivalent to (\ref{Ta0}). 
Its differential $\dd \V$ gives a two-form $\Om^2 = \dd \V$:
\begin{eqnarray} 
\hspace*{-5mm} \Om^2 = \dd \V & = & \Om_a\, \dd b \wedge \dd c + \Om_b \,\dd c 
     \wedge \dd a + \Om_c\,\dd a \wedge \dd b  = \bOm_a   \cdot \Surf  \nonumber  \\
   & = & \om_x\, \dd y \wedge \dd z + \om_y\, \dd z \wedge \dd x
		+ \om_z\, \dd x \wedge \dd y = \bom \cdot \surf , \label{Om2sS}	
\end{eqnarray}
where $\nabla_a\times \bV_a= (\Om_a, \Om_b,\Om_c)=\bOm_a$, and  $\nabla\times \bv
= (\om_x, \om_y, \om_z)=\bom$ is the vorticity.  Thus, it is seen that $\bOm_a$
is the vorticity transformed to the $\ba$-space, and the 2-forms $\surf$ and $\Surf$ 
are  surface forms defined in the footnote\footnote{$\surf = (\dd x^2 \wedge \dd x^3,\, 
\dd x^3 \wedge \dd x^1,\,\dd x^1 \wedge \dd x^2)$, and $\Surf =(\dd a^2 \wedge 
\dd a^3,\, \dd a^3 \wedge \dd a^1,\,\dd a^1 \wedge \dd a^2)$.}.   
The equation (\ref{dtOm-a}) is transformed to the $\tau$-derivative  of the 2-form 
$\Om^2$, which is understood as the  Lie derivative: $\Ld_{\pl_\tau} \Om^2=0$.  
Its representation in the $(\bx,t)$ space is given as follows.

It is useful to define two {\it tangent  vectors} $W$ and $V$ (tangent to the flow
generated by $\bv$): 
\begin{equation}
W \equiv \pl_t + V,  \hskip10mm
V \equiv u \pl_x + v\pl_y + w\pl_z = v^k\pl_k = \bv\cdot \nabla,      \label{TV-VW}
\end{equation} 
where $V$ is regarded as a tangent vector dual to the 1-form 
$\V$ of (\ref{V1xx}).  The Lie derivative $\Ld_{\pl_\tau}$ can be defined by 
\begin{equation}
\Ld_{\pl_\tau} = \Ld_W = \Ld_{\pl_t + V} = \pl_t  + \Ld_{V},    \label{Lda-x}
\end{equation} 
according to the symbols of differential calculus and tangent vector, where 
$V=v^k\pl_k$ is also a differential operator along a streamline 
generated by the flow $(u,v,w)$.  By applying the Lie derivative $\Ld_{\pl_\tau}$ on 
the 2-form $\Om^2$ (Frankel 1997, Ch.4.2),  we obtain 
\begin{eqnarray}	
0= \Ld_{\pl_\tau} \Om^2 = \pl_t \Om^2 + \Ld_{V}\Om^2 & = & \pl_t \bom \cdot \surf 
	+ [\bv,\bom] \cdot \surf + (\nabla\cdot \bv) \bom \cdot \surf	\nonumber  \\
 = \ \Ld_W \Om^2 \hskip11mm  & = &  E_W[\bom] \cdot \surf,   	\label{LdOm}
\end{eqnarray}
where $[\bv,\bom]$ is the Lie bracket defined by $(\bv\cdot\nabla)\bom 
- (\bom\cdot\nabla)\bv$, and 
\begin{equation}
E_W[\bom] \equiv \pl_t \bom  + [\bv,\bom] + (\nabla\cdot \bv) \bom 
	= \pl_t \bom + \nabla \times (\bom \times \bv), 		\label{VEq}
\end{equation} 
where  $\nabla \cdot \bom=0$ is used. Thus, we obtain the vorticity equation
from (\ref{LdOm}):
\begin{equation}
 \pl_t \bom + \nabla \times (\bom \times \bv) = 0.		\label{VEq0}
\end{equation} 
Next, let us introduce a gauge vector-potential $\bA_a=(A_a,A_b,A_c)$ in the 
$\ba$-space,  and define  its 1-form $A^1$ by 
\begin{eqnarray}
A^1    & = & A_a\,\dd a + A_b\, \dd b + A_c\,\dd c \hskip5mm 
	(\mbox{written as} \ \ \bA_a \cdot \dd \ba)  \label{A1-aa} \\
        & = & A_x\,\dd x + A_y\, \dd y + A_z\,\dd z.  \nonumber \label{A1xx} 
\end{eqnarray}  
The exterior product of $A^1$ and $\Om^2$ results in a volume form $\dd^3\ba = \dd a \wedge 
(\dd b \wedge \dd c) =\dd b \wedge (\dd c \wedge \dd a) =\dd c \wedge 
(\dd a \wedge \dd b)$ multiplied by a scalar $\langle \bA_a, \,\bOm_a \rangle$:
\begin{equation}
A^1  \wedge  \Om^2 = (\bA_a \cdot \dd \ba) \wedge (\bOm_a   \cdot \Surf) 
	= \langle \bA_a, \,\bOm_a \rangle  \, \dd^3\ba \,,   \label{dvphi3} 
\end{equation}
where $\langle \bA_a, \,\bOm_a \rangle = A_a\Om_a + A_b\Om_b +A_c\Om_c$ is 
a scalar product of  $\bA_a$ and $\bOm_a$.

In the $\bx=(x,y,z)$ space, the same exterior product is given by 
\begin{equation}
A^1  \wedge \Om^2 = \langle \bA,\, \bom \rangle \,  \dd^3\bx \,,	
	\hskip15mm  \dd^3 \bx = \dd x \wedge \dd y \wedge \dd z. 	\label{dvphi4}
\end{equation} 
It is obvious  that the scalar product $\langle \bA,\, \bom \,\rangle$ 
is invariant under local rotational transformations in the $\bx$ space.

Thus, it is  proposed that a {\it possible} type of Lagrangian is of the form,
\[  \Lambda_A = - \int_{M} \langle \pl_\tau \bA_a,\, \bOm_a \,\rangle \,\dd^3\ba
  = - \int_{M} \Ld_{\pl_\tau}\left[\langle \bA_a,\,\bOm_a \,\rangle \,\dd^3\ba \right], 
\]   
where the time derivative $\pl_\tau$ was taken out of the inner-product since $\bOm_a$ 
is independent of $\tau$.  It follows that $\Lambda_A$ can be dropped in the action $I$ 
expressed with the  coordinate $\ba$ except the integrated terms depending on
the initial and final times of $\tau$-integration.  However, this Lagrangian becomes 
non-trivial when transformed to the physical-space coordinate $\bx$ because of the 
non-trivial Jacobian  of transformation.  It is shown later that  $\bA_a$  may be 
understood as a kind of vector potential associated with the gauge field $\bOm_a$.

\subsection{Eulerian coordinates}

From the equality of (\ref{dvphi3}) and  (\ref{dvphi4}),  we have 
$\langle \bA_a, \bOm_a \rangle \dd^3\ba = \langle \bA, \bom \rangle \dd^3\bx$. 
Using the definition (\ref{Lda-x}), 
\[   \Ld_{\pl_\tau}\, \left[ \langle \bA_a,\,\bOm_a \,\rangle \,\dd^3\ba  \right]
 = \Ld_{W}\,\left[ A^1 \wedge \Om^2 \right]= \left[\Ld_{W} \,A^1 \right] \wedge \Om^2, \]
where (\ref{LdOm}) was used.  With the same calculus as  in (\ref{dvphi4}),
 the right hand side is given by $\left[ \Ld_{W}\,A^1 \right] \wedge \Om^2 
= \langle \Ld_W \bA,\, \bom \rangle \,\dd^3\bx$, where the Lie derivative of 
a 1-form $A^1$ is defined by 
\begin{eqnarray}
\Ld_W A^1 & := &  (\pl_t A_i + v^k \pl_k A_i + A_k \pl_i v^k)\,\dd x^i
		   = \Psi_i \,\dd x^i, 		\label{LdXA1} \\
\Psi_i & := & (\Ld_W \bA)_i = \pl_t A_i + v^k \pl_k A_i + A_k \pl_i v^k    \label{Bi}
\end{eqnarray} 
(\eg \ Frankel (1997), \S4.2c)\footnote{Note that, if $f(t,\bx)$ is a scalar function, 
$\Ld_W f = \pl_t f + V f \equiv  \DD_t f$.}. Thus, the Lagrangian $\Lambda_A$ 
in the $\bx$ variables is 		
\begin{equation}
\Lambda_A = - \int_{M} \langle \Ld_W \bA,\, \bom \,\rangle \,\dd^3\bx 
	= - \int_{M} \langle \bPsi,\, \bom \,\rangle \,\dd^3\bx, \qquad 
	\bPsi \equiv \Ld_W \bA. 			\label{LA-x1}  
\end{equation}			
Integrating this by parts and using the relation $\bom = \nabla \times \bv$, we have 
\begin{equation}
\Lambda_A = - \int_{M}\langle \nabla \times \bPsi,\,\bv \,\rangle \,\dd^3\bx
	+ \mbox{Int}_S  = - \int_{M} \rho \,\langle \bb,\,\bv \,\rangle \,\dd^3\bx
	+ \mbox{Int}_S. \label{LA-x2}  
\end{equation}
where Int$_S$ denotes the integral over the surface $S$ bounding $M$.
This is the proposed Lagrangian $\Lambda_A$ in the physical space  $\bx$, where 
\begin{equation}
\rho\,\bb \equiv \nabla \times \bPsi = \nabla \times (\Ld_W \bA).  \label{LA-x3}  
\end{equation}
The Lagrangian $\Lambda_A$ of (\ref{LA-x1})  is  an integral of $\langle \Ld_W \bA,\, 
\bom \,\rangle \equiv \Ld_W A^1 [\bom]$. Given a vector field $Y$, the value 
$\Ld_W A^1[Y]$ taken on the vector $Y$ measures the derivative 
of  $A^1$ (as one moves along the trajectory of $W$, \ie the particle path)  evaluated 
on a vector field $Y$  frozen to the flow generated by $W$. 

Using (\ref{Bi}) in (\ref{LA-x1}) and carrying out integration by parts, we obtain 
\begin{equation}
\Lambda_A =  \int_{M} \langle \bA,\, E_W[\bom] \,\rangle \,\dd^3\bx,    \label{LA-x4}  
\end{equation}
with $E_W[\bom]$ defined in (\ref{VEq}), where the integrated terms are dropped. 
Here,  we redefine $\Lambda_A$ by this equation instead of (\ref{LA-x1}).  Hence,
the expressions (\ref{LA-x1}) and (\ref{LA-x2}) must be modified by additional terms 
of surface integral over $S$, which are to be expressed symbolically with Int$_S$. 
This Lagrangian (\ref{LA-x4}) provides a source term of non-vanishing helicity. 
An example of such a flow will be given in  the last section \ref{8C}.
	
\section{Corrections to the previous Variation \label{S7}}

The Lagrangian $\Lambda_A$ of (\ref{LA-x2}) should be included in the total Lagrangian 
to describe rotational flows of an ideal fluid.  Accordingly, the Lagrangian density 
of the total Lagrangian (excluding surface terms) is defined by 
\begin{eqnarray}
L_{\rm T} = L_{\rm T}[\bv, \rho, s, \phi, \psi, \bb] & \equiv &
  \half\,\rho\,\langle\,\bv, \bv\,\rangle - \rho\,\eps(\rho,s) 
	-\rho\, (\pl_t+\bv \cdot\nabla)\phi		\nonumber  \\[1mm]
&&   \hspace*{-5mm}   -\rho s\, (\pl_t+\bv \cdot\nabla)\psi 
	-\rho\,\langle \,\bb,\, \bv\,\rangle.  	\label{Ld-ursps}
\end{eqnarray}
As before (Sec.~\ref{3B}), we take variations of the field variables $\bv$, \ $\rho$, 
\ $s$ and potentials $\phi$ and $\psi$. In addition, we must now include the variation 
of the new variable $\bA$.  Independent variations are taken for 
those variables. Substituting the varied variables \ $\bv+\del \bv$,\ $\rho +\del\rho$, \ 
$s +\del s$, \ $\phi +\del \phi$, \ $\psi +\del \psi$ and $\bA + \del \bA$ into 
$L_{\rm T}[\bv, \rho, s, \phi, \psi,\bA]$ and writing its variation as $\del L_{\rm T}$, 
we obtain 
\begin{eqnarray}
\del L_{\rm T} & = & \hskip4mm \delta \bv \cdot \,\rho (\bv - \nabla \phi
  -  s\,\nabla \psi - \bw \,)  - \del s\ \rho\ \DD_t\psi \label{du-ds} \\
&& \hspace*{-5mm} +\ \delta \rho\ (\half u^2 - h - \DD_t \phi 
		- s\,\DD_t \psi - \bv\cdot\bb \,) 	
+ \langle \delta \bA, \ E_{W}\bom \rangle	\nonumber \\
&& \hspace*{-5mm}	+ \hskip10mm \cdots \cdots 	\nonumber	\label{d-LdR}
\end{eqnarray} 
where remaining terms denoted by $\cdots \cdots$ are omitted since they are the 
same as (\ref{d-Ld}).  The new term $\bw$ is defined by
\begin{eqnarray}
\frac{\del\Lambda_A }{\del v^k}\,\del v^k & = & - \rho\,\langle \bw,\,\del \bv\,
	\rangle, 		\label{dLA-v} \\  
\bw & = &(w_i),\hskip10mm w_i= (\pl/\pl v^i)\langle \,\bv,\, \bb\,\rangle. \label{w-RS}
\end{eqnarray}
Thus, the variational principle, $\delta I =0$ for independent 
variations of $\delta \bv$, $\delta \rho$, $\del \bA$, \etc\, results in
\begin{eqnarray}
\delta \bv &:& \hskip5mm \bv -\nabla\,\phi- s\,\nabla \psi - \bw =0 \,, \label{dvel-r} \\
\delta \rho &:& \hskip5mm \half v^2 -h - \DD_t \phi - s\,\DD_t \psi 
		- \bv\cdot \bb = 0\,, \label{drho-b} \\
\delta \bA &:& \hskip5mm  E_{W}\bom =0\,,  \label{dA-om} \\
	&&  \hskip10mm \cdots \cdots \hskip10mm \cdots \cdots	 \nonumber 	
\end{eqnarray}
where the omitted terms  $\cdots \cdots$ are the same as before.  Thus, the  continuity 
equation (\ref{dphi}) and entropy equation (\ref{dpsi1}) are obtained as before.
The equation (\ref{dvel-r}) gives an expression for the velocity $\bv$:
\begin{equation} 
\bv = \nabla\, \phi + s\, \nabla \psi + \bw \,. 	\label{vel-a}
\end{equation}
The vorticity $\bom = \nabla \times \bv$  is represented by
\begin{equation} 
\bom = \nabla s \times \nabla \psi + \nabla \times \bw\,. 	\label{vor-a}
\end{equation}
The second term $\nabla \times \bw$ is a new term leading to  non-vanishing 
vorticity in a homentropic fluid of uniform $s$.  
 The  equation (\ref{dA-om}) is the same as (\ref{VEq0}) by (\ref{VEq}).

\vskip2mm
\noindent
{\it Potential part and rotational part}: \par 

The potential part of the present problem is described by (\ref{drho-b}),  and the 
rotational part  by (\ref{VEq0}).  In order to show that this is consisitent with 
the traditional formulation, suppose that the vorticity equation 
(\ref{dA-om}) is solved and a velocity field $\bv(\bx,t)$ is found.  
In view of the relation $\bom=\nabla \times \bv$, the velocity $\bv$ should satisfy 
the following:  
\begin{equation}
\pl_t \bv + \bom \times \bv = \nabla\,H',	 \label{EE-9a}
\end{equation}  
for a scalar function $H'(\bx,t)$.  Obviously, taking the curl of this equation reduces 
to (\ref{VEq0}).  

On the other hand, according to Sec.\ref{4B}, we have the Euler's equation 
of motion (\ref{EEq}),  because the new Lagrangian $\Lambda_A$ has no effect in  
the total Lagrangian of $\ba$ space owing to its construction, and hence generates 
no new term.  Using the identity (\ref{VI-S63}), this reduces to the following:
\begin{equation}
\pl_t \bv + \bom \times \bv = -\nabla(\half v^2 +h).	 \label{EE-9b}
\end{equation}  
Defining $H'$ by $H'=\pl_t \phi + s\,\pl_t \psi + \bv \cdot (\bb-\bw)$, 
the equation (\ref{drho-b}) is
\begin{eqnarray}
 0 & = & \half v^2 -h - \DD_t \phi - s\,\DD_t \psi - \bv\cdot \bb 
	= \half v^2 -h 	 - \pl_t \phi				\nonumber \\
   &  &  - s\,\pl_t \psi -\bv \cdot (\nabla \phi + s \nabla\psi + \bb) 	
		= - \half v^2 -h - H',   \label{drho8}
\end{eqnarray}
by using  (\ref{vel-a}).  In fact,  it can be shown that both expressions (\ref{EE-9a})
and (\ref{EE-9b}) are equivalent with this $H'$, as follows.  First we note 
the following identities:
\begin{eqnarray}
(\nabla \times \bw) \times \bv & = & (\bv\cdot \nabla)\bw 
	+ w^k\nabla v^k  - \nabla(\bv\cdot \bw), \label{id1} \\
w^k \nabla v^k - (\bv\cdot\nabla) \nabla \phi & = & \nabla(\half v^2 
	- (\bv\cdot \nabla) \phi) - s \pl_k\psi \nabla v^k .	\label{id2}
\end{eqnarray}
The second is an identity with $\bv$ given by (\ref{vel-a}).
Using these, one can verify the following:
\[   \pl_t \bw + (\nabla \times \bw) \times \bv +\nabla [\bv\cdot(\bw -\bb)]
	= \pl_t \bv + (\bv \cdot \nabla ) \bv +\nabla h.	\]
The right hand side vanishes by Eq.(\ref{EEq}).  Therefore, we obtain 
\begin{equation}
\pl_t \bw + (\nabla \times \bw) \times \bv = - \nabla [\bv\cdot(\bw -\bb)]. \label{Eq-w}
\end{equation}  
It is not difficult to show that the equation (\ref{EE-9a}) reduces to this equation for 
the above $H'$ and the total velocity $\bv$ of (\ref{vel-a}), where the total vorticity 
$\bom$ is given by (\ref{vor-a}).   The equation (\ref{Eq-w}) is a new equation to be 
satisfied by the rotational component $\bw$.

If $s=s_0$ (uniform), then $\bv=\nabla \Phi +\bw$ ($\Phi=\phi+s_0\psi$),
and $\bom=\nabla \times \bw$.  Subtracting (\ref{Eq-w}) from (\ref{EE-9b}),
we obtain $\nabla[\half v^2 +h +\pl_t \Phi + \bv \cdot (\bb-\bw)]=0$, 
which is equivalent to (\ref{drho8}).

 Thus there is no contradiction between the traditional formulation and the 
present one. New aspects of the present formulation are described in the next section,
in addition to the expression (\ref{Eq-w}) for the rotational component $\bw$.

\section{Significance  of  $\Lambda_A$ \label{S8} }

We consider here the significance  of the Lagrangian $\Lambda_A$ associated with 
rotation symmetry and its outcome derived from the variational principle.

\subsection{Source of helicity \label{8A} }

Let us recall that $\rho\,\bb$ is defined by (\ref{LA-x3}).  Using (\ref{Bi}), 
the expression of $\bPsi$ is rewritten as
\[ \bPsi = \pl_t A + v^k \pl_k A  + A_k \nabla v^k  
	= \pl_t A + (\nabla \times A)\times \bv + \nabla( A_kv^k) .  \]
Therefore, defining a vector $\bB$ with $\bB=\nabla \times A$, we have
\begin{equation}  
\rho\,\bb = 
\nabla \times \bPsi= \pl_t\bB +\nabla \times (\bB \times \bv) = E_W[\bB]. \label{rbb}
\end{equation}
Variation of $\Lambda_A$ of (\ref{LA-x2}) with respect to a variation $\del \bv$ 
is given by 
\[ \del \Lambda_A = - \int_{M}  \left[ \langle \nabla \times \bPsi,\,\del \bv \rangle
   +  \langle \nabla \times \del \bPsi,\,\bv \rangle\right] \,\dd^3\bx,  \]
where $\nabla \times \del \bPsi = \nabla \times (\bB \times \del \bv)$ from (\ref{rbb})
for the variation $\del \bv$ on keeping $\bA$ (hence $\bB$) fixed.
Therefore, we have 
\[ \del \Lambda_A = - \int_{M}  \Big\langle \big( \nabla \times \bPsi 
	+ \bom \times \bB \big),\,\del \bv \Big\rangle  \,\dd^3\bx.	\]
This leads to the following expression for $\bw$   defined by (\ref{dLA-v}):
\begin{equation}  
\rho \bw =\nabla \times \bPsi +\bom \times \bB =E_W[\bB]+ \bom \times \bB, \label{bw}
\end{equation}
where $E_W[\bB] \equiv \pl_t\bB + \nabla \times (\bB \times \bv)$ from (\ref{VEq}).

For flows of a homentropic fluid with constant entropy $s_0$, the velocity is 
given by (\ref{vel-a}): $\bv = \nabla\, \Phi + \bw$ ($\Phi= \phi + s_0 \psi$). 
Therefore, the helicity is 
\begin{equation}  
 H =  \int_V \ \bom \cdot \bv \ \dd^3\bx = \int_V \ \bom \cdot \bw \ \dd^3\bx 
	= \int_V \ \rho^{-1}\,  \bom \cdot E_W[\bB] \, \dd^3\bx,  
\label{Hlcty}
\end{equation}
since $\nabla\cdot \bom=0$. Thus it is found that the term $E_W[\bB]$ generates 
the helicity. 

\subsection{Uniqueness of transformation   \label{8B}}

Transformation from the  Lagrangian $\ba$ space to Eulerian $\bx(\ba)$ space is determined 
locally by nine components of the matrix $\pl x^k/\pl a^l$. However, in the previous 
solution considered in Sec.\ref{4B},  we had three relations (\ref{Web-Tr}) between 
the $\bx$-velocity $\bv=(X_\tau, Y_\tau, Z_\tau)$ and the $\ba$-velocity $(V_a, V_b,V_c)$, 
and another three relations (\ref{EqM-a-2}) between the $\bx$-acceleration 
$\A=(X_{\tau\tau}, Y_{\tau\tau}, Z_{\tau\tau})$ and the $\ba$-acceleration 
$(A_a, A_b,A_c)$. These six relations are not sufficient to determine the nine matrix 
elements $\pl x^k/\pl a^l$.  But now, we are equipped with a new structure of rotational 
symmetry which is expected to resolve  this indefiniteness. 

In fact, the remaining three conditions are found from the equation (\ref{Om2sS}) 
connecting the $\bx$-vorticity $\bom=(\om_x, \om_y, \om_z)$ and the $\ba$-vorticity 
$\bOm_a(\ba) =(\Om_a, \Om_b,\Om_c)$.  From (\ref{Om2sS}), these are
\begin{equation}  
\Om_a = \om_x\,(\pl_b y\,\pl_c z - \pl_c y\,\pl_b z) + \om_y\,(\pl_b z\,\pl_c x 
  - \pl_c z\,\pl_b x) + \om_z\,(\pl_b x\,\pl_c y - \pl_c x\,\pl_b y), \label{omx-Oma}
\end{equation}
with other two equations for $\Om_b$ and $\Om_c$  determined cyclically.

There are three vectors (velocity, acceleration and vorticity) determined by 
evolution equations subject to initial conditions in each space of $\bx$ and $\ba$ 
coordinates.  Transformation relations of the three vectors suffice to determine
the nine matrix elemets $\pl x^k/\pl a^l$ locally. Thus, the transformation between the  
Lagrangian $\ba$ space and Eulerian $\bx(\ba)$ space is determined uniquely.
In this sense, the equation of the vorticity is essential for the uniqueness
of the transformation between Lagrangian and Eulerian coordinates.

\subsection{Two linking vortices (an example) \label{8C}}

As an example of flows of non-zero helicity, we consider a particular flow field 
induced by tangling of two vortices, \ie a rectilinear line-vortex L linked with 
a vortex ring R of infinitesimal vortex-cores, whose cross-sections  are both  
circular and strengths are $\ga_L$ and $\ga_R$ respectively.  The helicity $H$
is immediately found as $2\ga_L \ga_R$ (see below), hence it is non-zero.  
Therefore, this flow field cannot be given a representation of a  Clebsch-type form.

We consider a flow of an incompressible fluid of density $\rho_0$ in  a cylindrical 
coordinate frame $(x, r, \phi)$. Suppose that we have a rectilinear line-vortex L 
of strength $\ga_L (>0)$ coinciding with the $x$-axis (axis of cylindrical symmetry) 
and a vortex ring R of strength $\ga_R (>0)$  which coincides temporarily with 
a  circle of radius $R$ in the plane $x=0$ centered at the origin.   By symmetry 
consideration, it is readily seen for this initial configuration that the vortex ring 
$R$ simply translates parallel to the $x$ axis with a constant velocity $U$ (say) 
without change of its  form and its radius (apart from rotaional motion around the 
$x$ axis without changing its form), while the rectilinear vortex $L$ stays at 
the same position without change. With respect to this coordinate frame, the velocity 
field (at the initial instant) of the rectilinear vortex L is represented by a vector 
potential $\Psi_L=(F(r), 0,0)$, while that of the vortex ring R is given by  
$\Psi_R =(0,\,0, \,G(x,r))$.  Explicit expressions of velocity are $\bv_L = 
\nabla \times \Psi_L$ and $\bv_R = \nabla \times \Psi_R$ respectively, and 
\begin{eqnarray*}
\bv_L & = &  (0,\,0, \, - \pl_r F),  \hskip10mm
F = - \frac{\ga_L}{2\pi} \, \log\,r, 		\label{VL1}	\\
\bv_R & = &  ( r^{-1} \pl_r(rG),\, - \pl_x G, \,0),  \hskip4mm  \\
 && \hspace*{15mm} G(x,r) = \frac{\ga_R}{4\pi}\,R \int_0^{2\pi} 
	\frac{\cos\phi \ \dd \phi}{\sqrt{x^2 + r^2 - 2rR \cos \phi +R^2}}. 
\end{eqnarray*}
Their vorticities are 
\begin{equation}
\bom_L =  (\ga_L\,\del(y)\del(z), \,0,\,0), \hskip10mm
\bom_R =  (0,\,0,\, \ga_R\,\del(r-R)\del(x)),         \label{Om-LR}
\end{equation}
where $\del$ is the delta function, and $(y,z)$ denotes the cartesian coordinates 
 in the plane perpendicular to the $x$-axis.  Hence, the helicity is given by
\[ H = \int_{\R^3} \left[ \bv_L \cdot \bom_R + \bv_R \cdot \bom_L \right]\, \dd^3\bx. \]
By elementary calculus, we find that the integral of each term gives the same value 
$\ga_L \ga_R$.  Thus we obtain  non-zero helicity $H= 2\ga_L \ga_R$ (Moffatt 1969).

The potential $\bB=\nabla \times A$ can be determined in the following way.  In 
reference to the coordinate frame moving with the vortex ring R, the velocity 
field is steady (and the vorticity too).  In this frame,  a uniform velocity
$(-U,0,0)$ is added, and total velocity is given by
\begin{equation}
\bw  = \bv_L + \bv_R + (-U,0,0) = \nabla \times \chi \equiv \bv(\bx) =(u,v,w),
\label{v-chi}
\end{equation} 
where $\chi = (F(r)-\half Ur, 0, G(x,r))=\chi(\bx)$, independent of $t$. 
In the equation (\ref{bw}): $\rho_0 \bv =  E_W[\bB] + \bom \times \bB$, the vorticity 
$\bom$ is nonzero only on the rectilinear line L and the circular curve R, 
and vanishes ($\bom=0$) at all other points. At points neither on L nor on R, we have
the equation:
\begin{equation}
\rho_0 \bv = E_W[\bB], \quad E_W[\bB]= \pl_t\bB + \nabla \times(\bB \times \bv). 
\label{B-LR}
\end{equation}
Let us seek a general solution to this equation for a given velocity field $\bv(\bx)$.
Suppose that we have a particle-path solution $\bx=\bX(\ba,t)$ satisfying
$\pl_t\bX(\ba,t)=\bv(\bX)$, where $\DD_t\ba=0$ for $\ba=\ba(\bx,t)$ and $\ba(\bx,0)=\bx$.
Next, we introduce a particular field  $\bB_*(\ba,t)$ 
defined by
\[   \bB_*(\ba,t) \equiv \Big(\bB_*(\ba,0) \cdot \frac{\pl}{\pl \ba}\Big) \bX(\ba,t), \]
which satisfies the equation: $E_W[\bB_*]= \pl_t\bB_* +\nabla \times(\bB_* \times \bv)=0$ 
(Cauchy's solution) for an arbitrary initial field $\bB_*(\bx,0)$.
A general solution to the equation (\ref{B-LR}) is given by 
\begin{equation}
\bB(\bx,t) = \bB_*(\ba,t) + \bB_0(\bx),    \label{gB-LR}
\end{equation} 
where $\bB_0(\bx)$ is a time-independent field and a particular solution to (\ref{B-LR}).
In fact, substituting (\ref{gB-LR}) into (\ref{B-LR}), we have $\rho_0 \nabla 
\times \chi = \nabla \times(\bB_0 \times \bv)$, in view of (\ref{v-chi}) and 
$E_W[\bB_*]= 0$.  Thus, we obtain
\[    \bB_0 \times \bv =  \nabla \psi + \rho_0  \chi \ \ (\equiv \bQ),   \]
where $\psi$ is a scalar function satisfying  $\bv\cdot \bQ=
(\bv\cdot\nabla)\psi+ \rho_0\bv\cdot \chi=0$.  Then, the solution $\bB_0$ is given by 
\[    \bB_0 = (0,\, -Q_\phi, Q_r)/u, \qquad \bQ=(Q_x, Q_r, Q_\phi).  \]

\section{Summary and discussion}

Following the scenario of the gauge principle of field theory, it is found that 
the variational principle of fluid motions can be reformulated successfully in terms 
of a covariant derivative and  Lagrangians defined appropriately.  The Lagrangians 
are determined such that a gauge invariance is satisfied under local Galilean 
transformations including the rotational symmetry.  The fluid  material 
is characterized thermodynamically in terms of mass density and entropy density.  
In addition, local rotation (\ie vorticity) of the material is also taken into account 
by a newly-introduced Lagrangian $\Lambda_A$.  A gauge-covariant derivative $\nabla_t$ 
was  defined in the present formulation including the rotational symmetry (it was 
$\DD_t$ up to \S 5 for the translation symmetry alone). Galilean invariance  
requires that the covariant derivative should be the {\it convective} time derivative 
following the motion of the background material, \ie the so-called Lagrange derivative. 
The covariant derivative is an essential building block of the gauge theory. 
Applying $\nabla_t$ to the position coordinate $\bx(\ba,t)$ of a fluid particle $\ba$, 
we obtain the particle velocity $\bv$. It is found that the vorticity $\bom$ is in 
fact a gauge field associated with the rotational symmetry.  Lastly, an example of 
flow with non-vanishing helicity is presented together with a general solution of the 
vector potential for the rotational part.

The variational principle (\ie the action principle) in the space of particle 
coordinate $\ba$ results in the Euler-Lagrange equation.  This yields the Noether thorem, 
\ie conservation equations of energy-momentum. The equations thus obtained 
are the Lagrangian form of the equation of motion and energy equation, which are 
transformed to  Euler's equation of motion and the conservation equation of 
total energy $\half v^2 + \eps$ in  Eulerian space.

The variational principle with respect to Eulerian space coordinate $\bx$ results 
in (i) the expression of velocity $\bv$ in terms of gauge potentials, (ii) an integral
describing a potential part of the velocity, (iii) the equation of continuity,
(iv) the entropy equation.  In addition to these results obtained in the 
previous case of translation symmetry alone, new results in the present formulation are 
as follows.  (v) The velocity $\bv$ includes a new rotational term $\bw$,
(vi) the vorticity equation (\ref{dA-om}) is deduced from the variational principle 
and (vii) the equation (\ref{Eq-w}) is derived for the rotational component $\bw$. 
It is noteworthy that the new Lagrangian $\Lambda_A$ of (\ref{LA-x4})  
yields a source term for the helicity defined by (\ref{Hlcty}).

Transformation from the Lagrangian $\ba$ space to Eulerian $\bx$ is determined locally 
by nine components of the matrix $\pl x^k/\pl a^l$. However, in the solution obtained 
previously (Kambe 2007) without taking account of rotational symmetry, we had only six 
relations for the velocity (\ref{Web-Tr}) and acceleration (\ref{EqM-a-2}) including 
the matrix element $\pl x^k/\pl a^l$. Both of the equations are  invariant under 
rotational transformations of a displacement vector $\Del \bX=(\Del X, \Del Y, \Del Z)$, 
so that the vector $\Del\bX$ is not uniquely determined.  In the improved formulation 
taking account of the rotational symmetry with three additional equations from 
(\ref{Om2sS}),  there are three vectors (velocity, acceleration and 
vorticity) determined by evolution equations subject to initial conditions in each 
space of $\bx$  and $\ba$.  Transformation relations of the three vectors suffice 
to determine the nine matrix elements $\pl x^k/\pl a^l$ locally. Thus, the transformation 
between the  Lagrangian $\ba$ space and Eulerian $\bx(\ba)$ space is determined 
uniquely.  In this sense, the equation  (\ref{omx-Oma}) for the vorticity is essential to
the uniqueness of the transformation between Lagrangian and Eulerian coordinates.
  
Accordingly, the present variational formulation is self-consistent and comprehensively 
describes flows of an ideal fluid.

\section*{Acknowledgment}

The author is grateful to Professor Antti Niemi for discussions on the Chern-Simons
term of the gauge theory, and also deeply appreciates the careful comments by 
Professor Alex Craik  on the original manuscript, which have led to significant 
improvement of the expressions. He is also much obliged to the referees 
and the editor for the comments to improve the manuscript.

\vskip10mm
\centerline{\bf Appendix: Background of the theory \label{A-Back}}
\appendix

This appendix reviews the background of the  theory  and  describes  the scenario 
of the gauge principle in a physical system.

\section{Gauge invariances \label{A1-GI}}

In the theory of   electromagnetism, it is well-known that there is an invariance 
under a {\it gauge transformation} of electromagnetic potentials consisting of 
a scalar potential $\phi$ and a vector potential $\bA$.  The electric field $\bE$ 
and magnetic field $\bB$ are represented as 	
$\bE= - \nabla \,\phi - (1/c)(\pl \bA/\pl t)$, and $\bB= \nabla \times \bA$,
where $c$ is the light velocity.  The idea is that the fields $\bE$ and $\bB$ are 
unchanged by the following transformation: $(\phi, \bA) \to (\phi', \bA')$, where 
$\phi'= \phi - c^{-1}\pl_t f$, $\bA'= \bA + \nabla\,f$ with $f(\bx, t)$ being 
an arbitrary differentiable scalar function of position $\bx$ and time $t$. 

Gauge invariance is also known in certain rotational flows of the Clebsch representation 
(Eckart 1960).  Its velocity field is represented as (Sec.\ref{3B}):
\[	\bv = \nabla \phi + s\,\nabla \psi,	\]
where $\phi,\ \psi$ and  $s$ are scalar functions, and $\psi$ satisfies the equation 
$\DD_t \psi=0$.  Let $\phi',\ \psi'$ and  $s'$ be a second set giving the same velocity 
field $\bv$, which implies the following: $s\,\nabla \psi - s'\,\nabla \psi' 
= \nabla (\phi' -\phi)$. This relation will hold true by the relation $\phi' -\phi 
= F(\psi,\,\psi')$,  if $s = \pl F/\pl \psi$ and $s' = - \pl F/\pl \psi'$. 
Therefore, $s$ and $\psi$ are determined only up to such a {\it contact} transformation,
and $\phi$ transforms by the addition of the generating function $F(\psi,\,\psi')$.
It can be shown  that the two sets of triplet give the identical 
flow field.   Thus, there exist some freedom in the expressions of physical 
fields in terms of potentials.  

\section{Related aspects in quantum mechanics and relativity theory \label{A2-QM}}

In {\it quantum mechanics}, Schr\"{o}dinger's equation for a charged particle of mass $m$   
and electromagnetic fields  are invariant with respect to a gauge transformation.  
This is as follows. In the {\it absence} of  electromagnetic fields, Schr\"{o}dinger's 
equation for a wave function $\psi$ of a free particle  $m$ is written as  
$S[\psi] \equiv i \hbar\, \pl_t \psi - \frac{1}{2m}\ p_k^2 \ \psi= 0$, where 
$p_k = -i \hbar\,\pl_k$ is the momentum operator ($\pl_k=\pl/\pl x^k$ for $k=1,2,3$).
In the {\it presence} of electromagnetic fields,  Schr\"{o}dinger's equation for a 
particle with an electric charge $e$ is obtained by the  transformation:
\begin{equation}	
	\pl_t  \to  \pl_t + (e/i\hbar) A_0, \hskip8mm 
	\pl_k  \to  \pl_k + (e/i\hbar c) A_k, \hskip4mm (k=1,2,3),  \label{S12-0}
\end{equation} 
where $(A_\mu)=(A_0,A_1,A_2,A_3)=(A_0,\,\bA)= (-\phi, \,\bA)$ is a four-vector potential, 
actually a covector denoted with lower indices.  A point in space-time is denoted by 
a four vector with upper indices, $(x^\mu)=(x^0,x^1,x^2,x^3)$  with $x^0=ct$. 
Thus,  we obtain  the  equation with electromagnetic fields:
\[  S_A [\psi] \ \equiv \ i \hbar\, \pl_t \psi - e\phi\,\psi - \frac{1}{2m}\, 
  \big[( -i \hbar)( \pl_k + \frac{e}{i\hbar c} A_k ) \big]^2 \ \psi =0. \]
From the four potential $A_\mu= (-\phi, \,\bA)$,  we obtain  $\bE$ and $\bB$.

Suppose that a wave function $\psi(x^\mu)$ satisfies the equation $\,S_{A}[\psi]=0$, 
and consider the following set of transformations of $\psi(x^\mu)$ and $A_\mu$:
\begin{eqnarray*} 
\psi'(x^\mu) & = &  \exp\left[ i\,\al(x^\mu)\right]\ \psi(x^\mu), \label{S82-psi}  \\
A_k & \to & A_k'=A_k+\pl_k \beta, \hskip8mm \phi \to \phi'=\phi - c^{-1} \pl_t \beta, 	
\end{eqnarray*}
where $\al = (e/\hbar c) \beta$.  Then, it is  shown readily 
that the transformed function $\psi'(x^\mu)$ satisfies the Schr\"{o}dinger equation 
$ S_{A'}[\psi']=0$.  This is the {\it gauge invariance} of the system of an electric 
charge in electromagnetic fields.  The system is said  to have a gauge symmetry.

A {\it relativistic} Lagrangian (\ie Lorentz-invariant) is defined for fluid flows by 
Soper (1976), and  the energy-momentum tensor $T^{\mu\nu}$ is derived from it,
where $\mu, \nu=0,1,2,3$ with 0 denoting the time component. 
In this relativistic formulation, the continuity equation and an energy equation of 
fluid flows are derived from the conservation of particle numbers and an equation of 
energy conservation $\pl_\nu\, T^{0\nu}=0$.   Expanding the energy density $T^{00}$ 
and energy flux $T^{0k}$ with respect to an infinitesimal ratio $\beta=v/c$ ($k=1,2,3$, 
where $c$ is the light velocity and $v$ a representative fluid velocity), the continuity 
equation is obtained from the lowest order terms (equivalent to the conservation 
of particle number),  while nonrelativistic energy conservation is obtained from 
the next order terms proportional to $\beta^2$.  The entropy equation is another 
form of the energy conservation.

\section{Brief scenario of gauge principle \label{A3-GP}}

In the gauge theory  of particle physics (Weinberg 1995; Frankel 1997),  a free-particle 
Lagrangian $\Lambda_{\rm free}[\psi]$ is defined first for the wave function 
$\psi(x^\mu)$ of a  particle with an electric charge.  Let us consider the following 
gauge transformation: $\psi \ \mapsto  \  e^{i\al}\,\psi$. If $\Lambda_{\rm free}$ 
is invariant under the transformation when $\al$  is a constant, it is said 
that $\Lambda_{\rm free}$ has a {\it global} gauge invariance.  In spite of this, 
it often happens that $\Lambda_{\rm free}$ is {\it not} invariant for a function 
$\al=\al(x)$, \ie \ $\Lambda_{\rm free}$ is not gauge-invariant {\it locally}.  
In this circumstance, it is  instrumental to introduce a new field in order 
to acquire local gauge invariance.   If the new field (a gauge field) is 
chosen appropriately, local gauge invariance can be recovered.  

In  the  section B,  the {\it local} gauge-invariance was acquired by replacing 
$\pl_\mu$ with 
\[	 \DD_\mu = \pl_\mu  + \A_\mu 	\]	
(see (\ref{S12-0})), where $\A_\mu =  (e/i \hbar c)  A_\mu$, and $A_\mu(x)$ is the 
electromagnetic potential (and  termed  a {\it connection form} in mathematics).
The operator $\DD_\mu $ is called the {\it covariant derivative}.

Thus, when the original Lagrangian is not locally gauge invariant, the {\it principle 
of local gauge invariance} requires a new gauge field to be introduced in order to 
acquire local gauge invariance, and the Lagrangian is to be altered  by replacing the 
partial derivative with the {\it covariant} derivative including the gauge field.  
This is the {\it Weyl's gauge principle}.  

In mathematical terms, suppose that we have a group $\G$ of transformations and 
an element $g(x) \in \G$ (for $x \in M$ with $M$ a space where $\psi$ is defined)
and that the wave function $\psi(x)$ is transformed as $\psi' = g(x) \psi$.  In the 
previous example,  $g(x)=e^{i\al(x)}$ and the group is $\G=U(1)$.  
Introducing the gauge field $A$ allows us to define a covariant derivative 
$\DD=\pl +A$ as a generalization of the partial derivative $\pl$ that transforms 
as $g\DD=g(\pl+ A) = (\pl'+ A')g$.  If we operates the right hand side on $\psi$, 
we obtain $(\pl'+ A')g \psi = (\pl'+ A')\,\psi'=\DD'\psi'$ where $\DD'=\pl' +A'$.  
Thus, we obtain $\DD'\psi'  = g\,\DD\psi$, showing that $\DD\psi$   transforms 
in the same way as $\psi$ itself.

In  dynamical systems which evolve with  the time  $t$, such as the present case of
fluid flows, replacement is to be made only for the $t$ derivative: $\pl_t \to  
\DD_t = \pl_t +  A(x)$.


\end{document}